\documentclass{article}
\usepackage[utf8]{inputenc}
\usepackage{color}

\usepackage[ruled,vlined]{algorithm2e}
\usepackage{amsmath}
\usepackage{amssymb}
\usepackage{caption}
\usepackage{graphicx}
\usepackage{natbib}
\usepackage{setspace}
\usepackage{subcaption}
\usepackage{tabularx}
\usepackage{verbatimbox} 
\usepackage[dvipsnames]{xcolor}

\DeclareMathOperator*{\argmax}{argmax}

\title{Fast Bayesian estimation of brain activation with cortical surface and subcortical fMRI data using EM}
\author{Daniel Spencer, David Bolin, Mary Beth Nebel, Amanda Mejia}
\date{\today}
\providecommand{\keywords}[1]
{
  \small	
  \textbf{\textit{Keywords---}} #1
}

\newcommand\Includegraphics[2][]{\addvbuffer[3pt 0pt]{\includegraphics[#1]{#2}}} 

\begin{document}

\maketitle

\begin{abstract}
    Analysis of brain imaging scans is critical to understanding the way the human brain functions, which can be leveraged to treat injuries and conditions that affect the quality of life for a significant portion of the human population. In particular, functional magnetic resonance imaging (fMRI) scans give detailed data on a living subject at high spatial and temporal resolutions. Due to the high cost involved in the collection of these scans, robust methods of analysis are of critical importance in order to produce meaningful inference. Bayesian methods in particular allow for the inclusion of expected behavior from prior study into an analysis, increasing the power of the results while circumventing problems that arise in classical analyses, including the effects of smoothing results and sensitivity to multiple comparison testing corrections. Recent development of a surface-based spatial Bayesian general linear model for cortical surface fMRI (cs-fMRI) data provides the desired power increase in task fMRI data using stochastic partial differential equation (SPDE) priors. This model relies on the computational efficiencies of the integrated nested Laplace approximation (INLA) to perform powerful analyses that have been validated to outperform classical analyses. In this article, we develop an exact Bayesian analysis method for the GLM, employing an expectation-maximization (EM) algorithm to find maximum a posteriori (MAP) estimates of task-based regressors on cs-fMRI and subcortical fMRI data while using minimal computational resources. Our proposed method is compared to the INLA implementation of the Bayesian GLM, as well as a classical GLM on simulated data. A validation of the method on data from the Human Connectome Project is also provided.
\end{abstract}


\section{Introduction}

Task-based functional magnetic resonance imaging (fMRI) is one of the best metrics currently available to allow researchers to understand the way that the human brain works \textit{in vivo}. Therefore, considerable effort has been made to wring out every last bit of inferential information from these data, a process that has spanned the thirty years since the technique was developed. Advancements in computer processing and storage technology have enabled more complex analyses of fMRI data, improving the ability of scientists to detect brain signals amidst the noise of measurement and physiological mechanics. Recent models are able to account for many sources of variance, accounting for spatial and temporal dependence within the data, though at the cost of computational efficiency.

Functional data from a magnetic resonance scanner come in the form of an array, in which the location within the array corresponds to the location of the measurement within the brain. The particular measurements that are taken are a unitless metric known as the blood oxygen level dependent, or BOLD measure. This measures the relative amount of blood in a given volume of the brain, with an understanding that the brain sends more blood to locations that are being used to process information \citep{lindquist2008statistical,poldrack2011handbook}. These small volumes within the brain to which these measurements correspond are called volumetric pixels, or voxels, each typically having a volume around $3mm^3$. Given that a scan will produce an image with hundreds of thousands of these measurements, care needs to be taken with any modeling decisions made in order to accurately account for different sources of variance. A commonly-used modeling technique is referred to in neuroscience as the \textit{general linear model}, which regresses the time series of measurements for each voxel individually against task covariates to determine the effect of the task on the movement of blood around the brain \citep{lazar2008statistical}. These regressions can be used to test whether a specific voxel has a significant association with the task covariates and is thus ``activated" by a task. As hundreds of thousands of statistical tests are performed, multiple testing corrections are necessary in order to reduce the chance of false positive results using techniques like the Bonferroni correction \citep{bonferroni1936teoria} and permutation testing \citep{nichols2002nonparametric}; or fixing the false discovery rate \citep{benjamini2000adaptive,benjamini2001control}. Brain functions have long been shown to be localized to specific parts of the brain \citep{broca1861remarks}, so accounting for spatial information within the array is highly important. To this end, various correction methods have been proposed to augment the GLM using various cluster-based methods \citep{poline1993analysis,poline1997combining,smith2009threshold}. However, such methods have been shown to produce inflated false positive rates, resulting in faulty inference \citep{eklund2016cluster,eklund2019cluster}. 

One solution to the problem of multiple testing corrections is to perform analysis using Bayesian methods and examine the posterior probability distribution to determine whether voxels are activated. Such techniques explicitly state the assumptions made in the model through the prior distributions, avoiding null hypothesis testing. \cite{zhang2016spatiotemporal} utilizes a Bayesian nonparametric framework to detect activations using both Markov chain Monte Carlo (MCMC) and variational Bayes (VB) methods for both single and multiple subjects, though even the VB method is rather expensive and requires some dimension reduction to perform analyses. \cite{siden2017fast} proposes a fast method using a spatial prior on volumetric single-subject data using both MCMC and VB methods that scales well. \cite{guhaniyogi2021bayesian} and \cite{spencer2020joint} use shrinkage priors and tensor decompositions to model volumetric task-based fMRI for both single and multiple-subject studies with promising results. However, all of these studies apply spatial priors on volumetric data using Euclidean distance to determine proximity, which does not represent the folded nature of the cerebral cortex.

Recently, \cite{mejia2020bayesian} proposes a surface-based spatial Bayesian (SBSB) GLM on cortical surface fMRI (cs-FMRI) data. Such data use a triangular mesh to find the geodesic distances between points on both the cortical surface and the subcortex, which appropriately accounts for the brain's folded structure. The SBSB GLM uses a stochastic partial differential equation (SPDE) prior to account for the dependencies in the Bayesian model, which first models individual subject data, which can then be combined in a principled manner to produce group-level inference. This model is validated to have attractive inferential properties by \cite{spencer2021spatial}, however it is computationally expensive to perform. 

We propose an efficient expectation-maximization (EM) algorithm to detect activation for task-based cortical surface fMRI data that drastically reduces evaluation time and memory needs. In addition, we extend the Bayesian GLM to subcortical data through increased memory efficiency in the EM algorithm. We compare the EM method to the INLA-based implementation of the SBSB GLM from \cite{mejia2020bayesian} using both simulated data and data from the Human Connectome Project (HCP) \citep{barch2013function} to show comparable results. We also implement the data modeling, including data preprocessing steps, within the open source R package \texttt{BayesfMRI}\footnote{https://github.com/mandymejia/BayesfMRI/tree/1.8.EM}.

This article will proceed with a description of the SBSB GLM and the EM implementation in section \ref{sec:methodology}. The methods are then compared in a study of simulated data in section \ref{sec:simulated}. Results from a study of the HCP data are shown in section \ref{sec:real}. We end with conclusions and discussion of future work in section \ref{sec:conclusion}.

\section{Methodology}
\label{sec:methodology}

Consider cs-fMRI or subcortical fMRI data from a scan, represented as $\mathbf{Y}_t \in \mathbb{R}^N$ for times $t = 1,\ldots,T$ at $N$ locations. These data are gathered while a subject completes $K$ different tasks as part of an experimental design. Data representing this design are represented as $\mathbf{X}_{t}^k \in \mathbb{R}^N$, which may not be identical across all locations after preprocessing (see section \ref{sec:preproc} for details). The $J$ nuisance regressors, accounting for unwanted effects such as motion and scanner drift, are represented through the notation $z_{t}^j$. Modeling brain activation for data in this form is done through the general linear model
\begin{align}
    \mathbf{Y}_t = \boldsymbol\mu + \sum_{k=1}^K \mathbf{X}_t^k\boldsymbol\beta^k + \sum_{j = 1}^J z_t^jb^j + \mathbf{e}_t \label{eq:GLM}
\end{align}
where $\boldsymbol\mu$ is the mean value, $\boldsymbol\beta^k$ is the regression coefficient for task $k$, $\mathbf{b}^j$ is the nuisance regression coefficient, and $\mathbf{e}_t$ is the error term. In our treatment of the model, the data are preprocessed to remove the mean effect $\boldsymbol\mu$. Please refer to section \ref{sec:preproc} for more details on data preprocessing. These data are then fitted with a model, with the main objective of performing inference about the parameters $\boldsymbol{\beta}^k$.

\subsection{Preprocessing} \label{sec:preproc}

Data preprocessing is a common practice in the modeling of neuroimaging data. It is often done with the intention of quickly removing variance stemming from known sources of error, such as movement and autocorrelation. We preprocess the data using functions within the \texttt{BayesfMRI} R software package, following the steps outlined below.

The values from the design matrix are preprocessed to account for the delay in task stimulus and physiological response through convolution with a haemodynamic response function (HRF), $h(t)$,
\begin{align*}
    x_{t,k} = \int_0^t x_{t,k}(\tau) h(t - \tau) d\tau.
\end{align*}
We use the canonical (double-gamma) HRF characterized as 
\begin{align}
    h(t) = \left(\frac{t}{a_1b_1}\right)^{a_1} e^{-(t - a_1b_1) / b_1} - c\left( \frac{t}{a_2b_2} \right) ^{a_2} e^{-(t - a_2b_2) / b_2}.
\end{align}
with values set according to default values given in the \texttt{neuRosim} package in R \citep{welvaert2011neurosim}. Specifically, the values used are $a_1 = 6$, $a_2 = 12$, $b_1 = b_2 = 0.9$, and $c = 0.35$. The values of the HRF-convolved design matrix $\mathbf{X}^k$ are then scaled by  dividing each column by its maximum value, then centering these values around 0. This is done to keep estimates of $\boldsymbol\beta_k$ comparable across different tasks. 

In order to facilitate spatial modeling given current computing memory capacity, the first step in preprocessing the response data is to downsample the data to a lower resolution, which is done for the HCP data using an interpolation method outlined in \cite{glasser2013minimal}. This downsampling presents a tradeoff between computational efficiency and spatial resolution in the inference, and needs to be considered thoroughly. This is examined in detail within \cite{spencer2021spatial}. After downsampling, the values from the response $\mathbf{Y}$ at each data location $v$ are centered and scaled using the function 
\begin{align}
    f(\mathbf{Y}_v) = 100 \times \frac{(\mathbf{Y}_v - \bar Y_v)}{\bar Y_v}, \label{eq:scaleY}
\end{align}
where $\mathbf{Y}_v$ is the cs-fMRI time series at location $v$, and $\bar Y_v$ is the average value at that data location across time. This transformation makes the fMRI time series interpretable as percent signal change, while also removing the need for mean value parameters, as in equation (\ref{eq:GLM}). 

Next, nuisance regression is performed to remove the effects of known confounding variables $z_t^j$ from the response data. This is done by regressing the centered and scaled response data against the nuisance variables, and then subtracting the estimated effects of the nuisance variables
$$ \tilde{\mathbf{Y}}_v = f(\mathbf{Y}_v) - \sum_{j=1}^J \mathbf{z}^j \hat{\mathbf{b}}^j, $$
where $f(\mathbf{Y}_v)$ is as defined in equation (\ref{eq:scaleY}), $\mathbf{z}^j \in \mathbb{R}^T$ is the value of the nuisance covariate across time, and $\hat{\mathbf{b}}^j$ is the regression estimate of the nuisance parameter found by regressing $f(\mathbf{Y}_v)$ against $\mathbf{Z} = (\mathbf{z}^1,\cdots,\mathbf{z}^J)$.

In order to remove temporal autocorrelation within the data, prewhitening is performed. This process first finds the residual values from a regression of the response ($\tilde{\mathbf{Y}}_v$) on the design matrix $\mathbf{X}_{v} \in \mathbb{R}^{T \times K}$, and then fits an AR(6) autoregressive model on these values using the method of solving the  Yule-Walker equations \citep{eshel2003yule}. The coefficients and the residual variance of the AR(6) model are then spatially smoothed using a Gaussian kernel with a full-width half-maximum (FWHM) of 6 mm. These are used to create a covariance matrix ($\mathbf{S}$) for the response at each location. The inverse of the square root of the covariance matrix ($\mathbf{D} = (\sqrt{\mathbf{S}})^{-1}$) is found using singular value decomposition. Finally, both the response data and the design matrix are premultiplied by $\mathbf{D}$ to produce the preprocessed response and task covariate data. 

After all of these preprocessing steps are applied, the data can be fit via a general linear model of the form
\begin{align}
    \mathbf{Y}_t = \sum_{k=1}^K \mathbf{X}_t^k\boldsymbol\beta_k + \mathbf{e}_t, \quad \mathbf{e}_t \sim \text{Normal}(\mathbf{0},\sigma^2\mathbf{I}). \label{eq:preprocessed_model}
\end{align}

\subsection{The Bayesian General Linear Model}

In order to facilitate computation, we change the model notation set forth in equation (\ref{eq:preprocessed_model}) in order to represent data across all locations and times at once. This altered notation is shown in equation \ref{eq:BayesianGLM}, in which the preprocessed response $\mathbf{Y} \in \mathbb{R}^{TN}$ at times $t = 1,\ldots,T$ and locations $i = 1,\ldots,N$ is explained by linear effects from each of $K$ tasks ($\mathbf{X}_k \in \mathbb{R}^{TN \times N}$) and an error term. 

\begin{align}
    \mathbf{Y}  = \sum_{k = 1}^K \mathbf{X}_k \boldsymbol{\beta}_k + \mathbf{e}, \quad \mathbf{e} \sim \text{Normal}(\mathbf{0},\mathbf{V}) \label{eq:BayesianGLM}
\end{align}

After prewhitening the data (please see section \ref{sec:preproc}), it is reasonable to assume that $\mathbf{V} = \sigma^2 \mathbf{I}_{TN}$, and the effects of the nuisance covariates are removed. This changed notation is used to show more clearly that the Bayesian GLM incorporates spatial dependence in the model itself, in contrast to the classical GLM, which does not include spatial dependence at the model-fitting stage.

\subsection{Spatial Process Prior}

In order to take full advantage of the cs-fMRI data and the improved interpretation of distance between two points, a triangular mesh with $n$ vertices imposed on the $N$ data locations (\cite{lindgren2011explicit}). A matrix $\boldsymbol{\Psi}_k \in \mathbb{R}^{N \times n}$ maps the original $N$ data locations to the $n$ mesh vertices. This process is done automatically to maximize the minimum interior angle of the triangles within the mesh, making transitions between small and large triangles more gradual, in the \texttt{R-INLA} package \citep{martins2013bayesian}. The prior on the coefficients for the task activation effects at the mesh vertices $\mathbf{w}_k$ is a special class of Gaussian Markov Random Field (GMRF) processes using stochastic partial differential equations (SPDEs) to discretize the continuous Mat\'ern covariance kernel to account for spatial similarities in the coefficients. Referred to as the SPDE prior \citep{lindgren2011explicit,bolin2013comparison}, a specific construction is imposed on the prior precision, shown in equation (\ref{eq:SPDEprior}). 
\begin{align}
    \boldsymbol{\beta}_k & = \boldsymbol{\Psi}_k \mathbf{w}_k \nonumber \\
    \mathbf{w}_k  \sim \text{Normal}(\mathbf{0},\mathbf{Q}_k^{-1}), &
    \quad \mathbf{Q}_k = \tau^2(\kappa^4 \mathbf{C} + \kappa^2 \mathbf{G} + \mathbf{G}\mathbf{C}^{-1} \mathbf{G}) \label{eq:SPDEprior} \\
    \kappa \sim \text{log-Normal}(\mu_\kappa,\sigma_\kappa^2), &
    \quad \tau \sim \text{log-Normal}(\mu_\tau,\sigma_\tau^2) \nonumber
\end{align}

This model is implemented as part of the R-INLA project\footnote{https://www.r-inla.org} \citep{rue2009approximate,thiago2013bayesian,lindgren2015bayesian}. In this parameterization, $\mathbf{C}$ is a fixed diagonal matrix describing the relative mesh vertex variances and $\mathbf{G}$ is a fixed sparse matrix describing the neighborhood structure of the $n$ mesh vertices, and both are calculated during model set-up.

\subsection{Full Vector Model Form}

For easier implementation, the model in equation (\ref{eq:BayesianGLM}) can be rewritten using the notation in equation (\ref{eq:SPDEprior}) to include all $k$ task covariates in matrix form:

\begin{align}
    \mathbf{Y} = \mathbf{X}\boldsymbol{\Psi}\mathbf{w} + \mathbf{e}, \quad \mathbf{e} \sim \text{Normal}(\mathbf{0},\mathbf{V}),
\end{align}

where $\mathbf{X} \in \mathbb{R}^{TN \times NK}$ is a column-bound matrix of the covariates, $\boldsymbol{\Psi} \in \mathbb{R}^{NK \times nK}$ is a block diagonal matrix projecting the covariate values onto the triangular mesh, and $\mathbf{w} \in \mathbb{R}^{nK \times 1}$ is the vector of covariates corresponding to the covariates on the mesh. 

\subsection{Expectation-Maximization Procedure}


The goal in using the expectation-maximization (EM) algorithm \citep{gelman2013bayesian} in this context is to find the mode of the posterior distribution of $\mathbf{w}$ by iteratively updating the parameter estimates for $\kappa_k$, $\tau_k$, and $\sigma^2$, and is shown in Algorithm \ref{alg:EM}. In addition, this method allows us to obtain an estimate of the posterior precision, which can be used in conjunction with the excursions method developed by \cite{bolin2015excursion,bolin2017quantifying,bolin2018calculating} to determine areas of activation in the latent field.

Here we choose to use the EM method as opposed to a more general variational Bayes (VB) method because inference for fMRI data centers around the posterior distribution for $\boldsymbol\beta_k$, and not the values for the parameters $\kappa^2$, $\tau^2$, and $\sigma^2$. Thus, assuming a point mass on the posterior values of these parameters does not have an appreciable effect on the inference for the parameters of interest, while allowing for rapid convergence to a solution comparable to that found using INLA while keeping memory usage to a minimum.

\begin{algorithm}[H]
\SetAlgoLined
\KwResult{Final estimate of the posterior mode of $\Theta = \{\kappa_1,\ldots,\kappa_K,\tau_1,\ldots,\tau_K,\sigma^2\}$}
Start with initial values for $\Theta$\;
\While{$|\Theta - \Theta_{\text{old}}| > \epsilon$}{
 Set $\Theta_{old} = \Theta$\;
 E-step: Update $E_{old}(\log p(\mathbf{w},\Theta_{old}|\mathbf{y}))$\;
 M-step: Set $\Theta = \underset{\Theta}{\mathrm{argmax}} \, \left. E_{old}(\log p(\mathbf{w},\Theta|\mathbf{y})) \right.$\;
 }
 Set $\mathbf{w} = E(\mathbf{w} | \Theta)$ \;
 \caption{Expectation-maximization method for finding the posterior mode of $\Theta = (\kappa,\tau,\sigma^2)$}
 \label{alg:EM}
\end{algorithm}

Following this algorithm, the estimates for $\Theta$ are found upon convergence, leading to the posterior full conditional distribution of $\mathbf{w}$.

\subsubsection{Expectation of the Log-Likelihood Density (E-step)}

Here we derive the expectation of the joint log likelihood density. The likelihood density can be found as 
\begin{align}
    p(\mathbf{y},\mathbf{w} | \Theta) & = p(\mathbf{y}|\mathbf{w},\sigma^2) p(\mathbf{w}| \kappa,\tau).
\end{align}
The expectation of the log likelihood density with respect to $\mathbf{w}$ is, therefore:
\begin{align}
    R(\Theta|\Theta_{\text{old}}) = E(\log p(\mathbf{y},\mathbf{w}|\Theta)) & \propto \int  \log p(\mathbf{y},\mathbf{w}| \Theta) p(\mathbf{w}| \mathbf{y}, \Theta_{\text{old}}) d\mathbf{w}, \label{eq:EoldFormula}
\end{align}
where $\mathbf{w} | \mathbf{y},\Theta \sim \text{Normal}\left(\boldsymbol{\mu}_{w|y}, \boldsymbol{\Sigma}_{w|y}\right)$, such that 
\begin{align}
    \boldsymbol{\Sigma}_{w|y} & = \left( \mathbf{Q} + \frac{1}{\sigma^2}\boldsymbol{\Psi}' \mathbf{X}' \mathbf{X} \boldsymbol{\Psi} \right)^{-1}, \\
    \boldsymbol{\mu}_{w|y} & = \frac{1}{\sigma^2}\boldsymbol{\Sigma}_{w|y} \boldsymbol{\Psi}'\mathbf{X}'\mathbf{y}.
\end{align}
We can compute $\boldsymbol{\mu}_{w|y}$ by setting $\mathbf{m} = \frac{1}{\sigma^2} \boldsymbol{\Psi}' \mathbf{X'y}$ and solving for $\mathbf{x}$ in the system of linear equations $\boldsymbol{\Sigma}_{w|y}^{-1}\mathbf{x} = \mathbf{m}$. In the interest of computational efficiency, several identities are leveraged to avoid expensive matrix inversions. The calculation of $\boldsymbol{\Sigma}_{w|y}$ must be avoided, as it requires the inversion of an $n \times n$ dense matrix, with $n$ between 1,000 and 30,000 in most practical applications. However, this computation is unnecessary for the calculation of the posterior means and the parameter estimates. In the next section, the MLE of $\sigma^2$ is found, and the optimization strategy for $\kappa$ and $\tau$ is described.

\subsubsection{Maximizing the Log Likelihood Density with Respect to $\kappa$, $\tau$, and $\sigma^2$}

Expanding and simplifying the expected log likelihood density in (\ref{eq:EoldFormula}) results in the expression in (\ref{eq:E_llik}), which must be optimized in order to estimate $\Theta = \{\kappa_1,\ldots,\kappa_K,\tau_1,\ldots,\tau_K,\sigma^2\}$.
\begin{align}
    R(\Theta|\hat{\Theta}) = E(\log p(\mathbf{y},\mathbf{w}|\Theta)) = R_1(\Theta|\hat{\Theta}^{(s)}) + R_2(\Theta|\hat{\Theta}^{(s)}) \label{eq:E_llik}
\end{align}

The expected log-likelihood can be further factorized in order to isolate the terms that involve $\sigma^2$ and find its MLE:
\begin{align}
    R_1 & (\Theta|\hat{\Theta}^{(s)}) \propto \nonumber \\ 
    & -\frac{TN}{2} \log(\sigma^2) - \frac{1}{2\sigma^2}\mathbf{y}'
    \mathbf{y} + \frac{1}{\sigma^2} \mathbf{y}'\mathbf{X}\boldsymbol{\Psi}\text{E}(\mathbf{w})  - \frac{1}{2\sigma^2} \text{Tr}(\boldsymbol{\Psi}' \mathbf{X}' \mathbf{X} \boldsymbol{\Psi} \text{E}(\mathbf{ww}')) \label{eq:R1} \\
    R_2 & (\Theta|\hat{\Theta}^{(s)}) \propto \frac{1}{2} \log |\mathbf{Q}| - \frac{1}{2} \text{Tr}(\mathbf{Q}\text{E}(\mathbf{ww}')), \label{eq:R2}
\end{align}
where $|\mathbf{A}|$ is the determinant of the matrix $\mathbf{A}$, and $\text{Tr}(\mathbf{A})$ is the trace of matrix $\mathbf{A}$. The posterior distribution for $\mathbf{w}$ has the moments $E_{w|y}(\mathbf{w}) = \boldsymbol{\mu}_{w|y}$ and $E_{w|y}(\mathbf{ww}') = \boldsymbol{\Sigma}_{w|y} + \boldsymbol{\mu}_{w|y}\boldsymbol{\mu}_{w|y}'$.  The MLE for $\sigma^2$ can be found through the maximization of $R_1(\Theta|\hat{\Theta})$ with respect to $\sigma^2$:
\begin{align*}
    \frac{\partial R_1}{\partial \sigma^2} & = -\frac{TN}{2\sigma^2} + \frac{1}{2(\sigma^2)^2}\mathbf{y'y} - \frac{1}{(\sigma^2)^2} \mathbf{y}'\mathbf{X}\boldsymbol{\Psi}\text{E}(\mathbf{w}) + \frac{1}{2(\sigma^2)^2}\text{Tr}(\boldsymbol{\Psi}' \mathbf{X}' \mathbf{X} \boldsymbol{\Psi} \text{E}(\mathbf{ww}')), \\
    \widehat{\sigma^2} & = \frac{1}{TN} \left[ \mathbf{y'y} - 2\mathbf{y'X}\boldsymbol{\Psi}E(\mathbf{w}) + \text{Tr}(\boldsymbol{\Psi}'\mathbf{X'X}\boldsymbol{\Psi} E(\mathbf{ww'}) \right] \label{eq:sigma2_hat}
\end{align*}

Next, values of $\tau$ and $\kappa$ must be found that maximize the log-likelihood. This is equivalent to optimizing equation (\ref{eq:R2}) with respect to $\kappa$ and $\tau$. However, given the multiplicative relationship between $\tau^2$ and $\kappa^2$, a reparameterization is needed to avoid identifiability issues in the maximization step when finding $\kappa^2$ and $\tau^2$.

\subsubsection{Reparameterization of the SPDE Precision Structure}

Due to difficulties with identifiability in the maximization step of the EM algorithm with the parameterization above, we consider an alternative parameterization in which 
\begin{align}
    \phi_k & = c_1(\kappa_k^2\tau_k^2)^{-1}, \quad c_1 = \frac{1}{4\pi}, \\
    \rightarrow \mathbf{Q}_k & = \frac{c_1}{\phi_k}\left( \kappa_k^2 \mathbf{C} + 2\mathbf{G} + \kappa_k^{-2} \mathbf{GC}^{-1}\mathbf{G} \right). 
\end{align}
With this parameterization, the form of the expected log-likelihood would remain the same as in (\ref{eq:E_llik}), (\ref{eq:R1}), (\ref{eq:R2}). Thus, the MLE for $\sigma^2$ remains the same as in (\ref{eq:sigma2_hat}). If (\ref{eq:R2}) is rewritten as 
\begin{align}
    R_2\left(\Theta | \hat{\Theta}^{(s)} \right) & =  \frac{1}{2} \log \left| \mathbf{Q}\right| - \frac{1}{2} \text{Tr} \left(  \mathbf{Q} \text{E}(\mathbf{ww}') \right) \\
    & = \frac{1}{2} \log\left( \prod_{k=1}^K \left|\frac{c_1}{\phi_k}\tilde{\mathbf{Q}}_k\right| \right) -\frac{1}{2}  \text{Tr}\left( \mathbf{Q} \text{E}(\mathbf{ww}') \right) \\
    & = \frac{n}{2} \sum_{k=1}^K \log \left( \frac{c_1}{\phi_k} \right) +\frac{1}{2} \sum_{k=1}^K \log( |\tilde{\mathbf{Q}}_k| ) -\frac{1}{2} \text{Tr}\left( \mathbf{Q} \text{E}(\mathbf{ww}') \right) \\ 
    & = \frac{nK}{2} \log (c_1) - \frac{n}{2} \sum_{k=1}^K \log (\phi_k) +\frac{1}{2} \sum_{k=1}^K\log( |\tilde{\mathbf{Q}}_k| ) - \frac{1}{2} \text{Tr}\left( \mathbf{Q} \text{E}(\mathbf{ww}') \right) \nonumber \\
    \mathbf{Q} & = \text{diagonal}\left(\frac{c_1}{\phi_1}\tilde{\mathbf{Q}}_1,\ldots,\frac{c_1}{\phi_K}\tilde{\mathbf{Q}}_K\right) \nonumber \\
    \tilde{\mathbf{Q}}_k & = \left( \kappa_k^2 \mathbf{C} + 2\mathbf{G} + \kappa_k^{-2} \mathbf{GC}^{-1}\mathbf{G} \right), \nonumber
\end{align}
then 
\begin{align}
    \frac{d}{d\phi_k} R_2 & = -\frac{n}{2\phi_k} + \frac{c_1}{2\phi_k^2} \text{Tr} (\tilde{\mathbf{Q}_k} \text{E}(\mathbf{w}_k\mathbf{w}_k')), \\
    \hat{\phi}_k & = \frac{c_1}{n} \text{Tr}(\tilde{\mathbf{Q}_k}\text{E}(\mathbf{w}_k\mathbf{w}_k')).
\end{align}
Using this reparameterization, the optimal value for $\kappa$ is found by maximizing the function in equation (\ref{eq:kappa_obj}).
\begin{align}
    R_2 \left( \kappa_k | \hat{\phi}_k, \hat{\Theta}^{(s)} \right) \propto \frac{1}{2}\log\left( |\tilde{\mathbf{Q}}_k| \right) - \frac{c_1}{2\hat{\phi}_k} \text{Tr}\left( \tilde{\mathbf{Q}}_k \text{E}(\mathbf{w}_k\mathbf{w}_k') \right) \label{eq:kappa_obj}
\end{align}
This allows for the estimates for $\phi_k$ and $\kappa_k$ to be found iteratively, avoiding identifiability present when finding the joint maximum. Additionally, these task-specific hyperparameters can be found in parallel across tasks, which allows for faster computation times.

\subsubsection{Maximizing Computational Efficiency}

Aside from resampling the data to a lower resolution, a number of steps are taken to decrease computational burden, both in terms of time and memory usage. First, since the left and right hemispheres of the cortical surface are physically independent, they are analyzed using the Bayesian GLM separately in parallel rather than together in order to decrease computation time. Second, the initial values for the hyperparameters are found by finding the classical GLM estimates of the task coefficients ($\boldsymbol{\beta}^k$) (please see section \ref{sec:classicalGLM}). Next, given these estimates, the maximum likelihood estimates of $\kappa_k$ and $\phi_k$ are found for $k = 1,\ldots, K$, as well as the maximum likelihood estimate for $\sigma^2$ these are found using a stepwise optimization algorithm (see \textbf{Appendix \ref{app:initial}}). Additionally, the initial values and the EM task-specific hyperparameters are found in parallel across tasks, reducing the computational load associated with analysis with more tasks. The linear algebra library and optimization is done by using functions within the \texttt{INLA} R package \citep{rue2009approximate,thiago2013bayesian,lindgren2015bayesian} using the PARDISO solver \citep{schenk2002two}. Finally, squared extrapolation methods are used to speed up convergence via the \texttt{SQUAREM} package in R \citep{du2020squarem}.

\subsection{Group-level analyses}

One of the benefits of using cortical surface data is the ability to reliably map to a standard, allowing for population-level inference about brain function. Combining data across subjects is done in a principled, straightforward manner in the work done by \cite{mejia2020bayesian}, which we will briefly outline. First, the summary statistics $\boldsymbol{\Phi}'\mathbf{X}_m'\mathbf{X}_m\boldsymbol{\Phi}$ and $\boldsymbol{\Phi}'\mathbf{X}_m'\mathbf{y}_m$ are pulled from each individual-subject analysis for subjects $m = 1,\ldots, M$. In the EM implementation, a group estimate for the hyperparameters $(\boldsymbol{\theta}_G)$ is found via a weighted average across the subject-level analyses. 

Next, the posterior mean and precision for $\boldsymbol{\theta}_m = (\kappa_{m,1},\ldots,\kappa_{m,K},\phi_{m,1},\ldots,\phi_{m,K},\sigma_m^2$ are combined using a weighted average for the mean and the sum for the precision. These values are used to draw a number of posterior samples of a group estimate $\boldsymbol{\theta}_G$ for the hyperparameters, which are then used to obtain a number of posterior samples of the group estimates $\boldsymbol{\beta}_{G}$ for the brain activity association. These samples are then used to perform inference on population-level associations and activations. 

\subsection{The Classical General Linear Model} 
\label{sec:classicalGLM}

In contrast to the Bayesian GLM, the classical GLM does not account for any spatial dependencies in the data in the modeling step itself. Each data location $v$ is modeled separately, resulting in the model
\begin{align}
    \mathbf{Y}_v & = \sum_{k=1}^K \beta_{v,k} \mathbf{x}_{v,k} + \mathbf{e}_v, \label{eq:classicalGLM}
\end{align}
assuming the data have been preprocessed to remove the effects of the nuisance regressors and eliminate the need to estimate a mean value. The notation represents the cs-fMRI or subcortical fMRI time series as $\mathbf{Y}_v \in \mathbb{R}^T$ and design matrix columns as $\mathbf{x}_{v,k}$ so that the linear models across all locations can be fit quickly and in parallel, if necessary. In general, this classical GLM has the advantage of being able to be fit quickly, even to full-resolution data. However, this advantage comes at the cost of ignoring the spatial nature of activations, which reduces model power significantly.

\section{Simulated Data Analysis}
\label{sec:simulated}

Data were simulated for cortical surface in R using a software package we developed called \texttt{brainSim}\footnote{https://github.com/danieladamspencer/brainSim/tree/1.0}. These simulations create brain surface data and design matrices with spatially-dependent coefficients and autoregressive error terms. The software also allows for a variable number of subjects, scanning sessions, and scanning runs, which allows for nested individual, session, and run variances from a ``true" global coefficient. 

\subsection{Single-subject analysis}

We begin with simulations for a single subject, simulating data under nine different conditions. These conditions are the combination of resampling resolution at $n = \{2,000; 5,000; 10,000\}$ cortical surface vertices per hemisphere, and setting the number of tasks to be $K = \{2,5,8\}$. Under these conditions, the length of the scan is set to $T = 300$, the maximum value for the spatial coefficient is set to 2. Under these conditions, the average value for the nonzero true coefficients is approximately 0.5 due to spatial smoothing in the data generation. The error term was set to have a variance of 1, and the error was assumed to be temporally independent.

\begin{figure}
    \centering
    \begin{subfigure}[b]{0.49\textwidth}
        \centering
        \includegraphics[width=\textwidth]{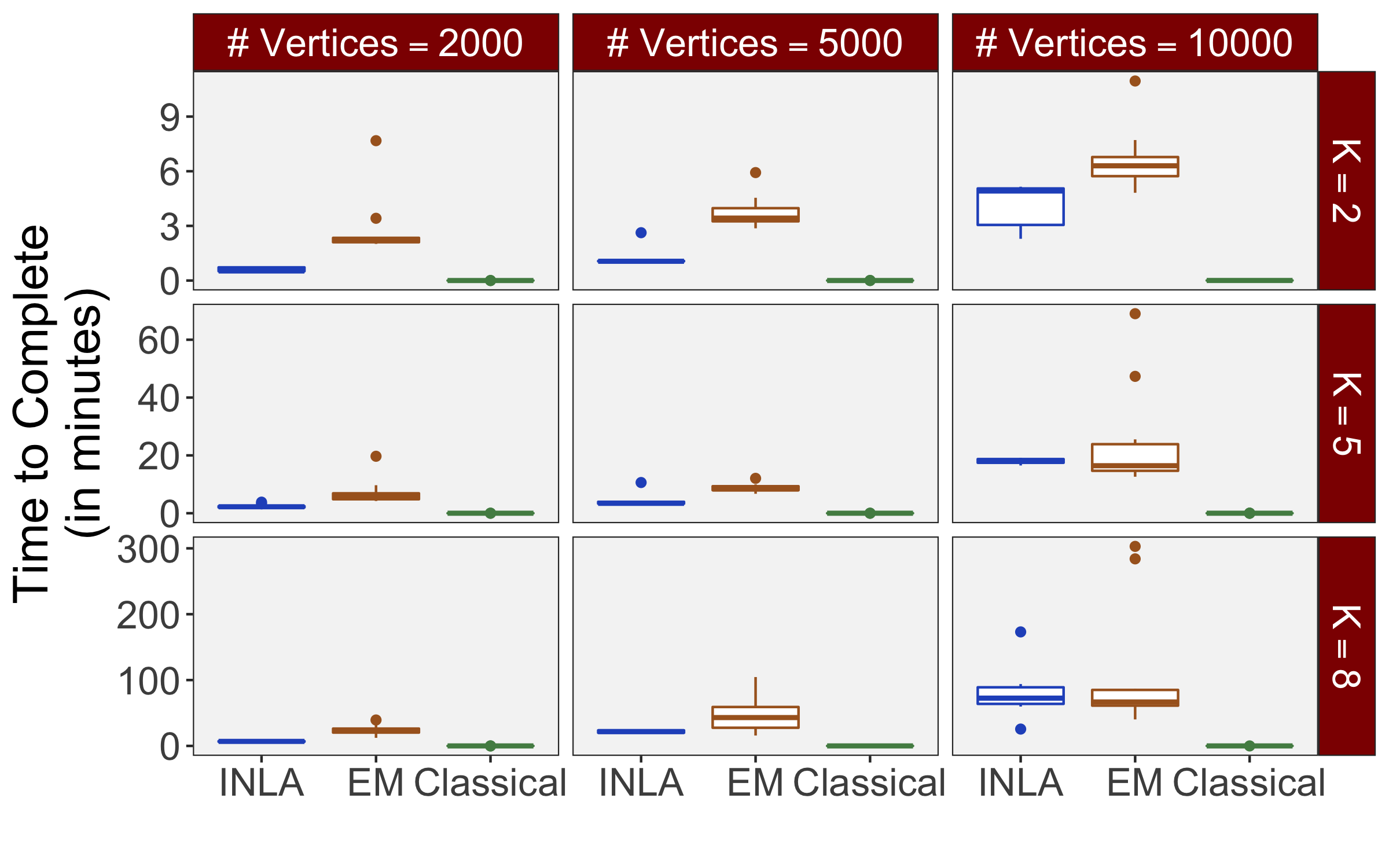}
        \caption{Time after preprocessing}
        \label{fig:time_sim}
    \end{subfigure}
    \hfill
    \begin{subfigure}[b]{0.49\textwidth}
        \centering
        \includegraphics[width=\textwidth]{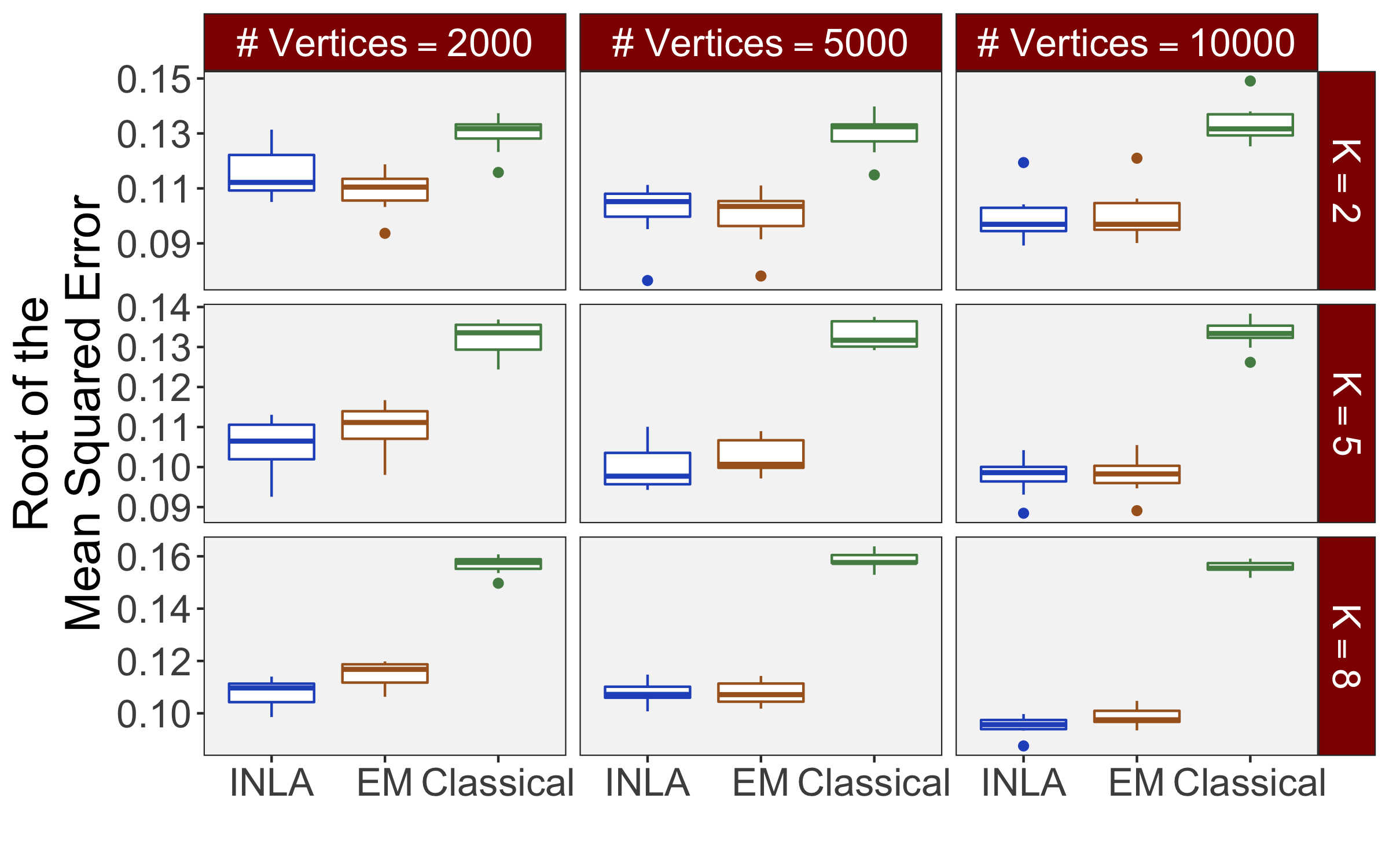}
        \caption{RMSE}
        \label{fig:rmse_sim}
    \end{subfigure}
    \caption{Performance comparison for the single-subject, simulated data under the four different simulation scenarios.}
\end{figure}

\begin{figure}
    \begin{subfigure}{0.49\textwidth}
        \begin{tabularx}{\textwidth}{c|c|c|}
    		\multicolumn{1}{c}{} & 
    		\multicolumn{1}{c}{\textbf{1 Run}} & 
    		\multicolumn{1}{c}{\textbf{2 Run}} \\ 
    		\cline{2-3} 
    		\rotatebox[origin=l]{90}{\quad \textbf{Truth} \qquad \,} &
    		\Includegraphics[width=.4\textwidth]{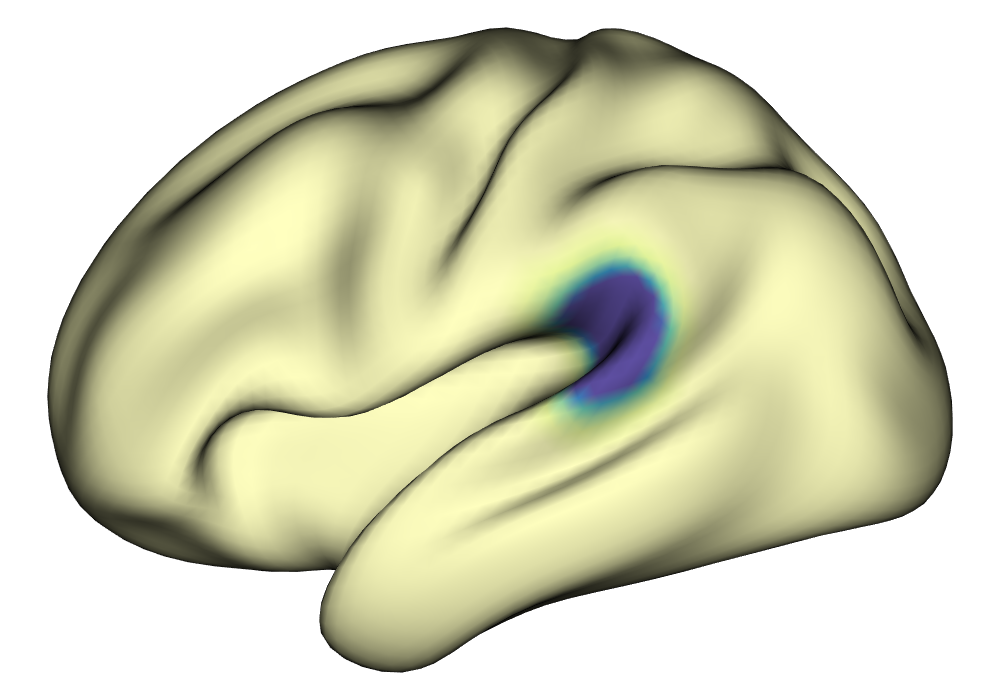} &
    		\Includegraphics[width=.4\textwidth]{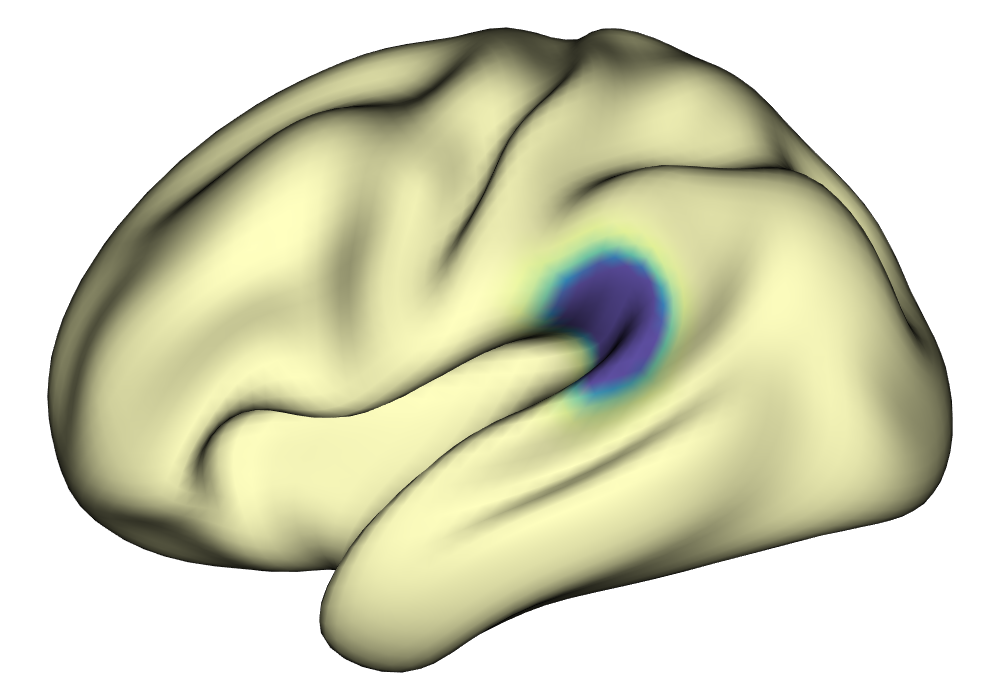} \\
    		\cline{2-3}
    		\rotatebox[origin=l]{90}{\quad \textbf{INLA} \qquad \,} &
    		\Includegraphics[width=.4\textwidth]{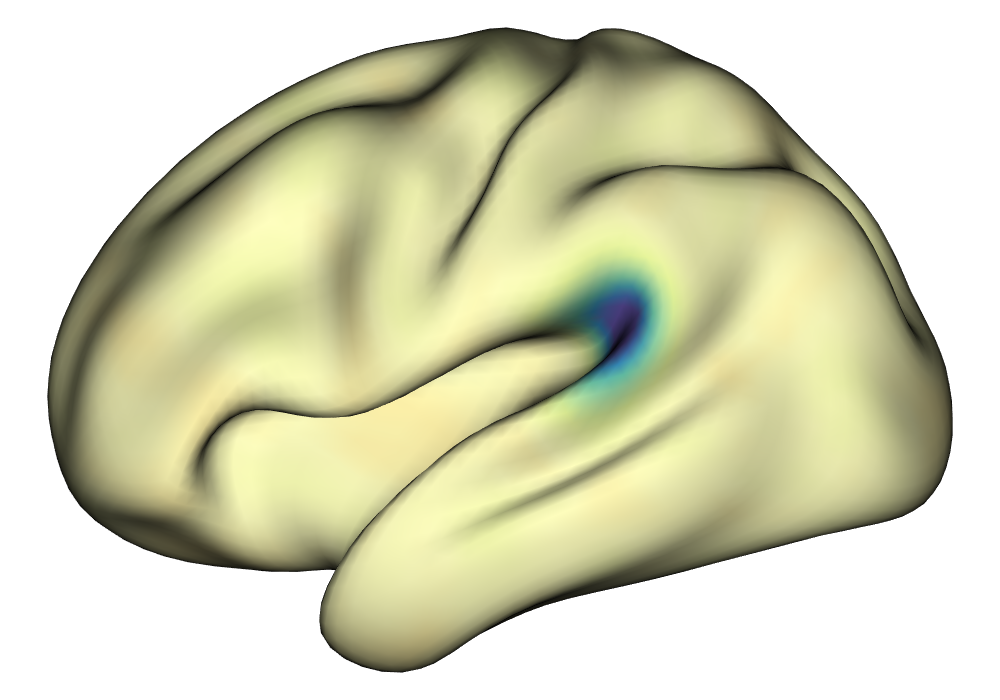} &
    		\Includegraphics[width=.4\textwidth]{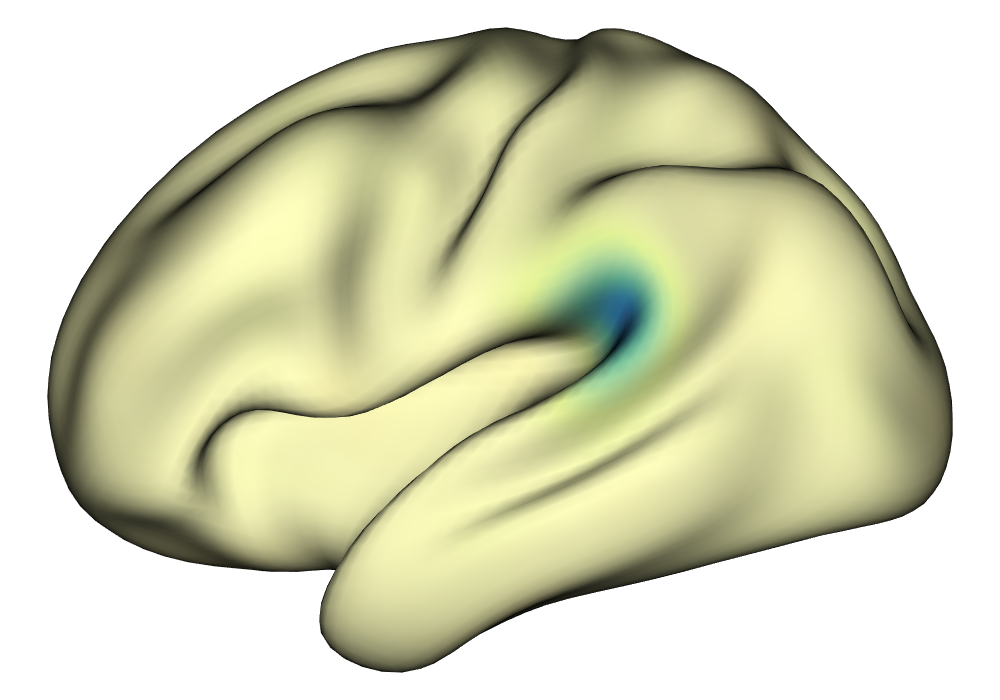} \\
    		\cline{2-3}
    		\rotatebox[origin=l]{90}{\quad \textbf{EM} \qquad \,} &
    		\Includegraphics[width=.4\textwidth]{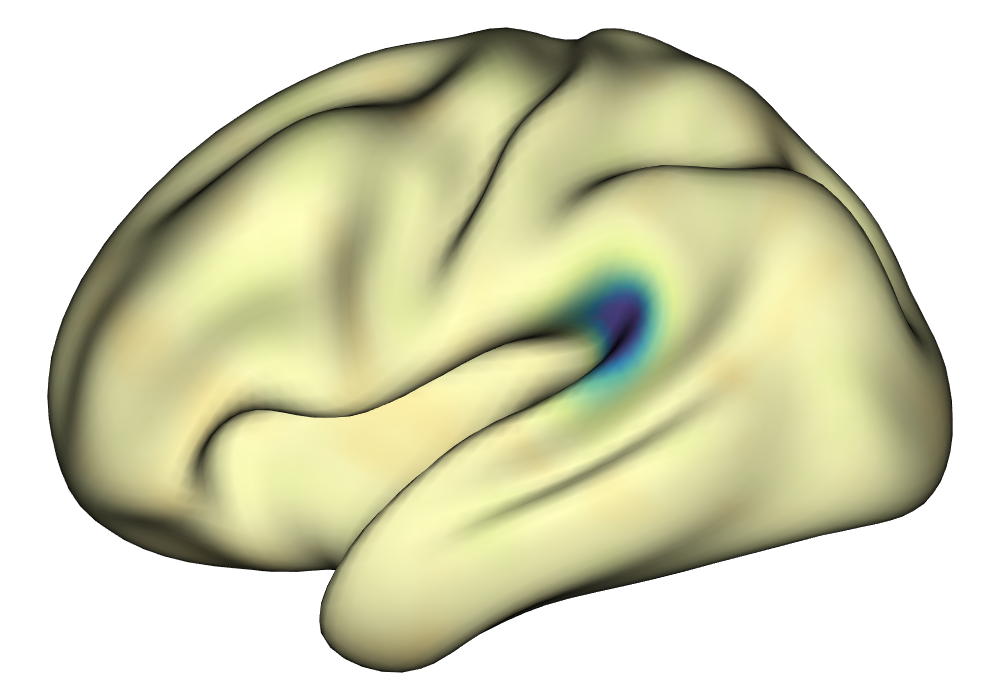} &
    		\Includegraphics[width=.4\textwidth]{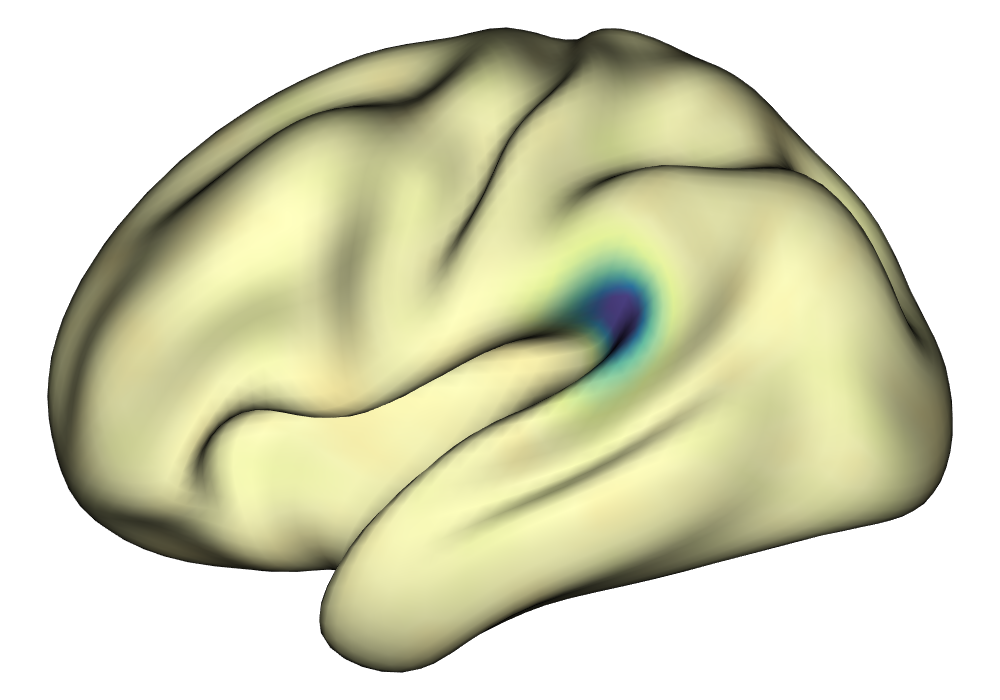} \\ 
    		\cline{2-3}
    		\multicolumn{1}{c}{\rotatebox[origin=l]{90}{\qquad}} & \multicolumn{2}{c}{\includegraphics[width=.6\textwidth]{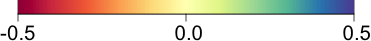}}
	    \end{tabularx}
	    \caption{Estimates}
	    \label{fig:sim_estimations}
    \end{subfigure}
    \begin{subfigure}{0.49\textwidth}
        \begin{tabularx}{\textwidth}{c|c|c|}
            \multicolumn{1}{c}{} & 
            \multicolumn{1}{c}{\textbf{1 Run}} & 
            \multicolumn{1}{c}{\textbf{2 Run}} \\ 
    		\cline{2-3} 
    		\rotatebox[origin=l]{90}{\quad \textbf{Truth} \qquad \,} &
    		\Includegraphics[width=.4\textwidth]{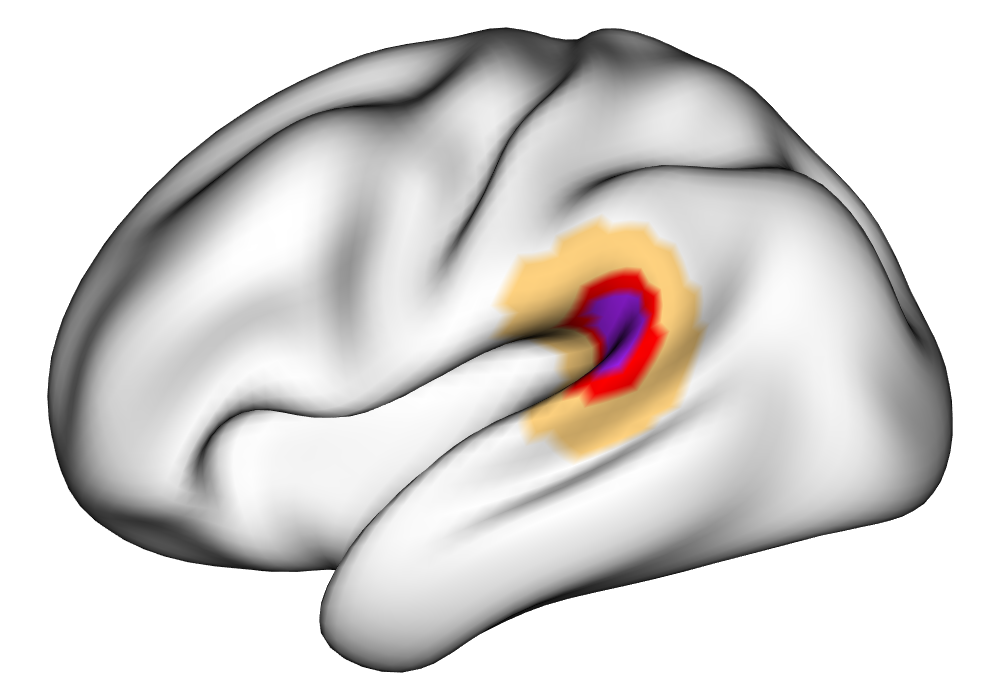} &
    		\Includegraphics[width=.4\textwidth]{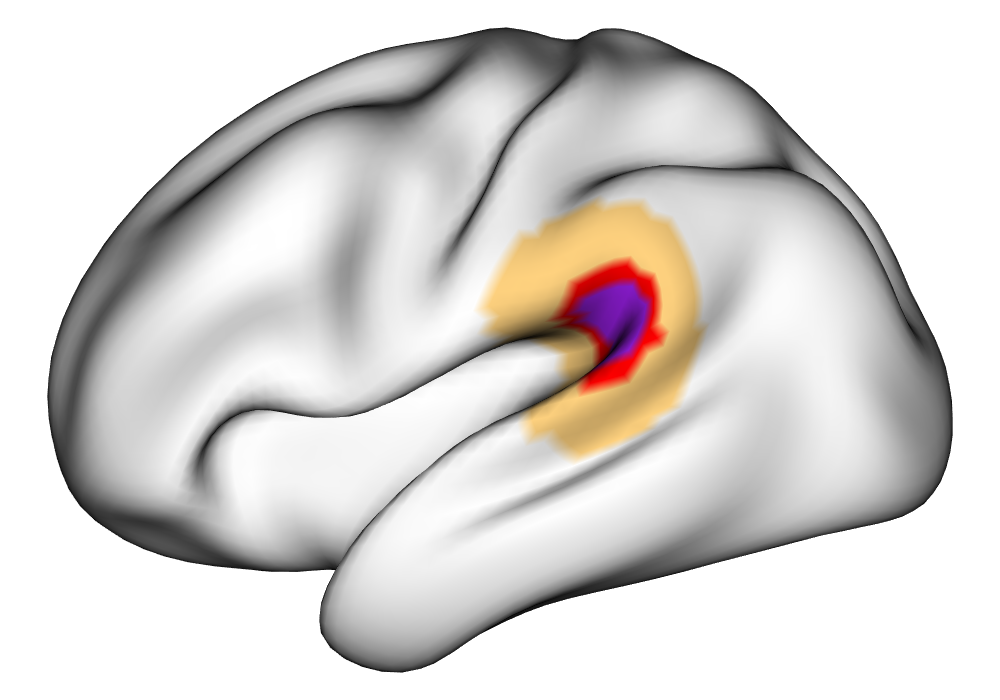} \\
    		\cline{2-3}
    		\rotatebox[origin=l]{90}{\quad \textbf{INLA} \qquad \,} &
    		\Includegraphics[width=.4\textwidth]{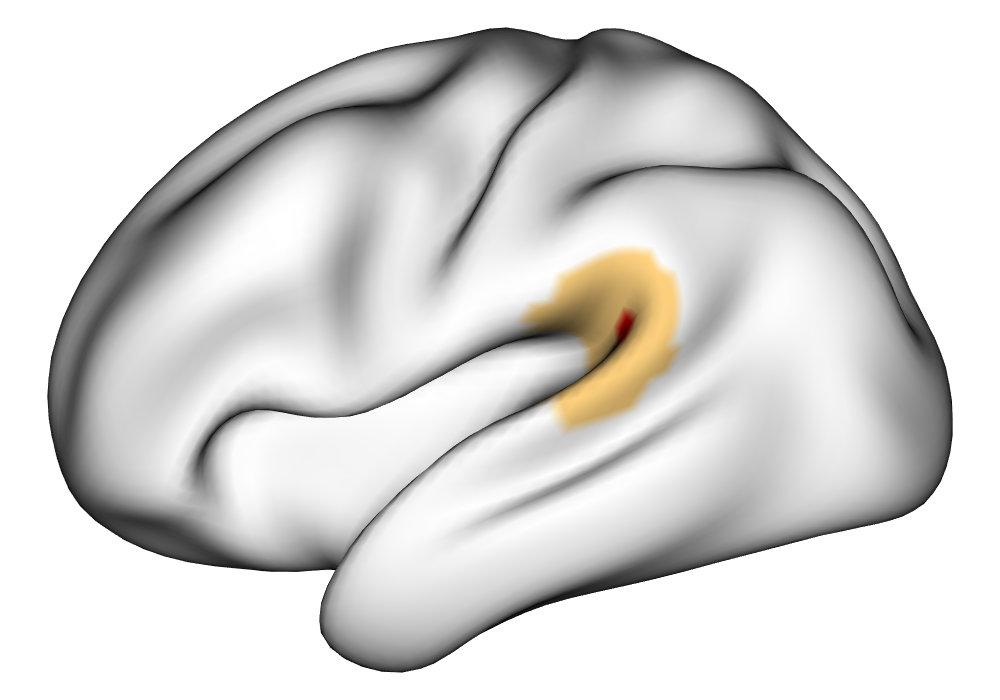} &
    		\Includegraphics[width=.4\textwidth]{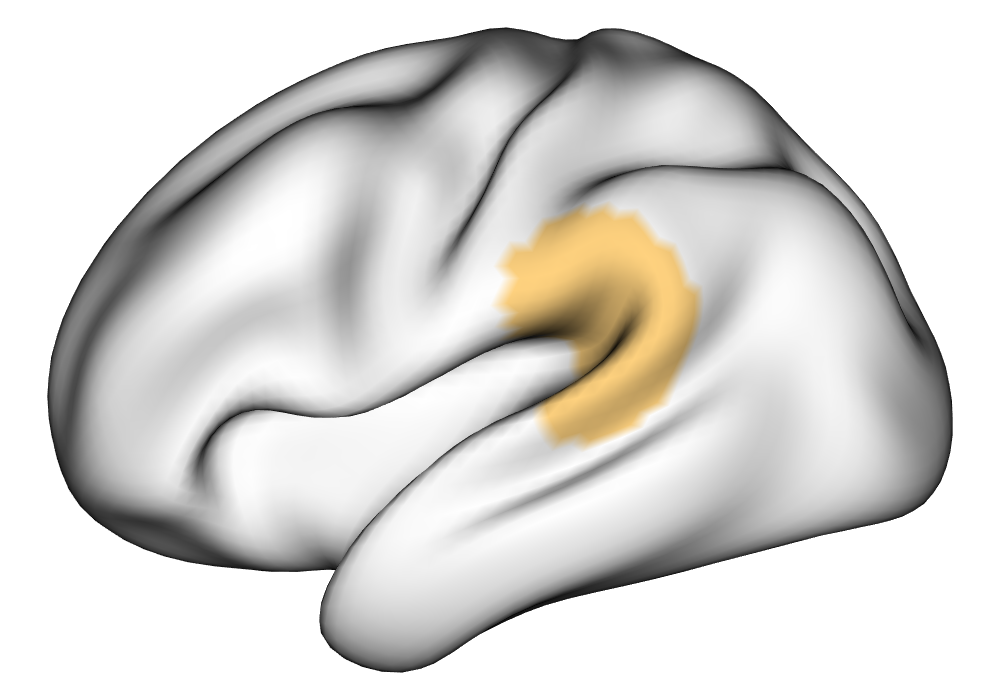} \\
    		\cline{2-3}
    		\rotatebox[origin=l]{90}{\quad \textbf{EM} \qquad \,} &
    		\Includegraphics[width=.4\textwidth]{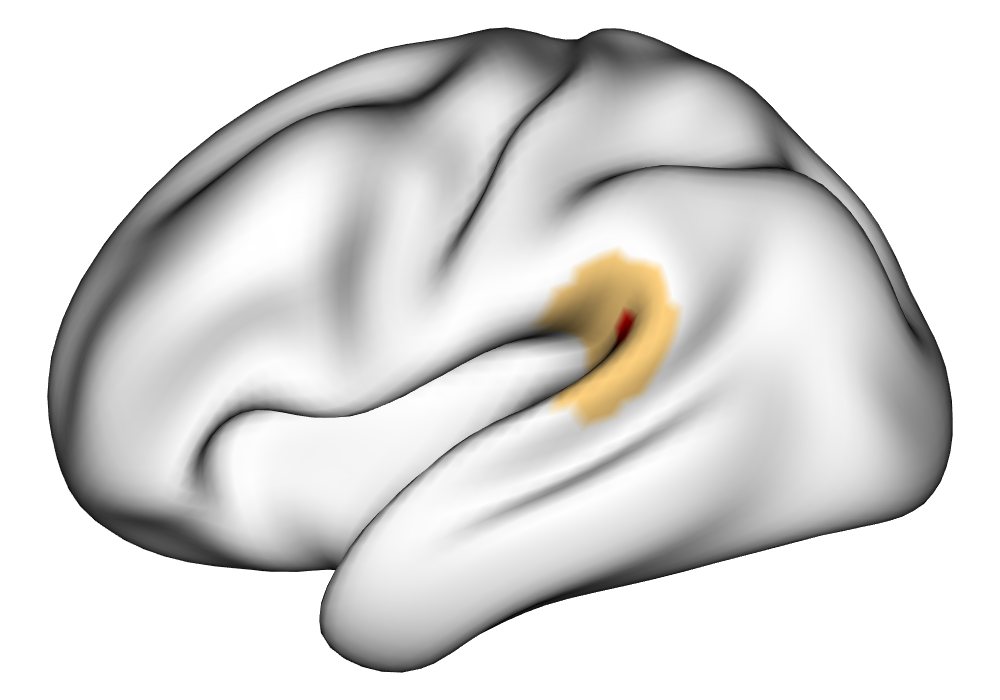} &
    		\Includegraphics[width=.4\textwidth]{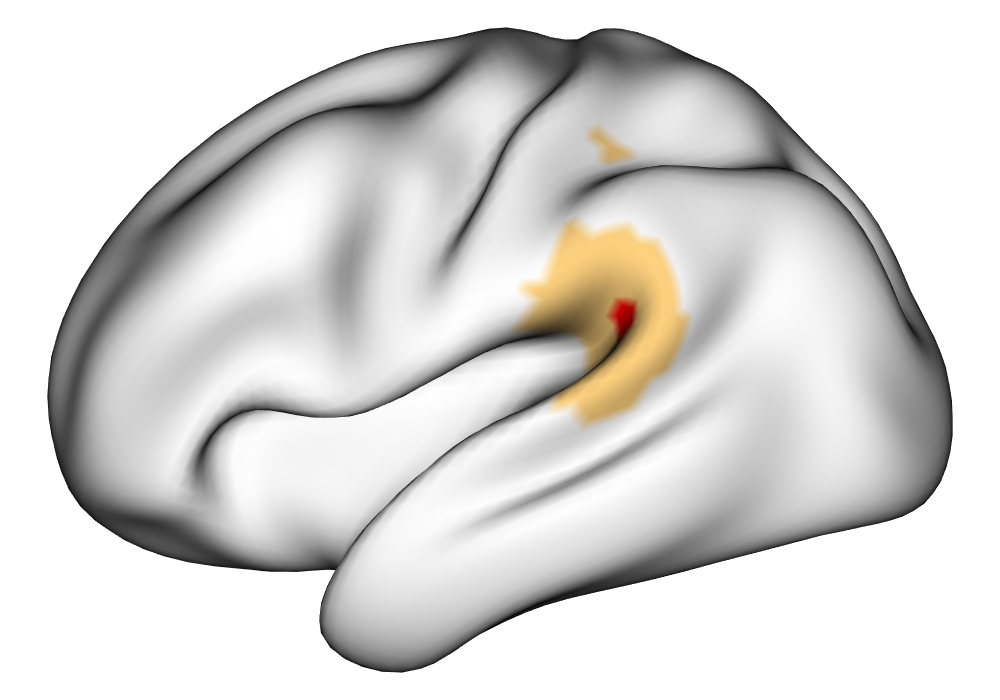} \\ 
    		\cline{2-3}
    		\multicolumn{1}{c}{\rotatebox[origin=l]{90}{\qquad}} & \multicolumn{2}{c}{$\gamma =$ \textcolor[HTML]{FFD27F}{$\blacksquare$} 0\% 
           \textcolor[HTML]{FF0000}{$\blacksquare$} 0.5\% 
           \textcolor[HTML]{A020F0}{$\blacksquare$} 1\%}
		\end{tabularx}
		\caption{Activations}
		\label{fig:sim_activations}
    \end{subfigure}
    \caption{Cortical surface coefficient estimates and activations in percent signal change for the INLA and EM implementations of the Bayesian GLM. True estimates and activations are shown for comparison. The activations are found using the excursions method, and are found as areas deemed jointly greater than threshold $\gamma$.}
    \label{fig:sim_est_and_act}
\end{figure}

Each of the nine conditions was used to generate 10 datasets, which were all fit to the classical GLM, and the Bayesian GLM using the INLA and EM implementations. The results were then compared in terms of the amount of time to perform the analysis after preprocessing (\textbf{Figure \ref{fig:time_sim}}) and in terms of their root mean squared error (RMSE) (\textbf{Figure \ref{fig:rmse_sim}}). These results show that the EM implementation is comparable in terms of computation time with the INLA implementation when the data are relatively small, as when the spatial resolution is set to 1000 vertices per hemisphere and there are only 2 covariates. However, increasing either the number of vertices or the number of covariates results in an increasing time advantage to the EM implementation of the Bayesian GLM. This is achieved while maintaining almost identical performance in terms of the RMSE. A visualization of an estimated and true coefficient surface for the sampling condition when $n = 5000$ and $K = 5$ can be seen in \textbf{Figure \ref{fig:sim_estimations}}.

Under the simulation condition with resolution $n = 5000$ and number of tasks $K = 5$, simulated data were generated for two runs in a single session to allow for sharing of the hyperparameters $(\kappa_1,\ldots,\kappa_5,\phi_1,\ldots,\phi_5,\sigma^2)$ across the two runs. The INLA and EM implementations of the Bayesian GLM were then used to analyze the first run by itself and the two runs combined. The effect estimates and activations illustrating the difference in the 1- and 2-run analyses can be seen in \textbf{Figure \ref{fig:sim_est_and_act}}. As there is a small run effect simulated in the generation of the data, the true values of the activation amplitude are not identical between the two runs. Therefore, the true coefficient nonzero region in the single run is not centered exactly in the same location as the true nonzero coefficient region across both runs. In terms of both the coefficient estimation and activation, the EM implementation performs very similarly to the INLA implementation, though there is a small region of spurious activation found by the EM implementation when both runs are analyzed together. This is likely due to the EM algorithm's underestimation of the posterior variance of $\boldsymbol{\beta}$, which stems from treating the hyperparameters $\Theta = \{\kappa_1,\ldots,\kappa_K,\tau_1,\ldots,\tau_K,\sigma^2\}$ as fixed rather than varying, as in the INLA implementation.

\subsection{Group analysis}

For population-level inference, it is important to see how well the model performs with simulated data in order to provide information about inferential performance with a known truth. In order to assess this ability, the \texttt{brainSim} R package was used to generate data for 10 subjects with a common true value, allowing for small spatial variation in the true values for each subject. The single-subject model was fit using the INLA and EM implementations of the Bayesian GLM as well as the classical GLM for all 10 subjects. 

\section{Analysis of HCP Data}
\label{sec:real}


The Human Connectome Project \citep{barch2013function} motor task and gambling test-retest data were analyzed using the classical GLM, the Bayesian GLM implemented using INLA, and the Bayesian GLM EM algorithm in order to compare the methods and their performance. These data were collected from 45 subjects across two scanning sessions, with two fMRI runs performed for each session. For our analysis here, we examine the first visit for 10 subjects in order to mimic smaller sample sizes seen in many imaging studies, as \cite{spencer2021spatial} shows the INLA implementation of the Bayesian GLM to be powerful in such small studies.

These data were obtained from the HCP after being preprocessed with the minimal preprocessing pipeline. This pipeline includes the projection of the volumetric blood oxygen level-dependent (BOLD) values to the cerebral cortex and the subcortical regions and registering these surfaces to a common surface template. The preprocessing also creates high-resolution surface meshes with 164,000 vertices using structural T1-weighted and T2-weighted MRI scans, which are resampled to 32,000 vertices per cortical surface hemisphere to match the resolution of the fMRI scans. In order to regularize the mapping process to the cortical surface, the fMRI timeseries were smoothed using a Gaussian smoothing kernel with a 2mm full-width half-maximum (FWHM). 

After these steps from the HCP minimal preprocessing pipeline, additional preprocessing as outlined in section \ref{sec:preproc} are applied in order to remove nuisance effects and temporal autocorrelation in the data, and make the parameter estimates of $\boldsymbol{\beta}_k$ interpretable in terms of percent signal change, given the unitless nature of fMRI BOLD measures. 

Following these preprocessing steps, the resulting data for both runs of the first scanning session was analyzed separately for each hemisphere and each subject using the INLA and EM implementations of the Bayesian GLM. Here we examine the estimates and activations detected by the model implementations to show consistent results using real data.

\begin{figure}
    \begin{subfigure}{0.49\textwidth}
        \begin{tabularx}{\textwidth}{c|c|c|}
            \multicolumn{1}{c}{} &
            \multicolumn{1}{c}{\textbf{INLA}} &
            \multicolumn{1}{c}{\textbf{EM}} \\
            \cline{2-3} 
    		\rotatebox[origin=l]{90}{\textbf{Subject A}} &
    		\Includegraphics[width=0.4\textwidth]{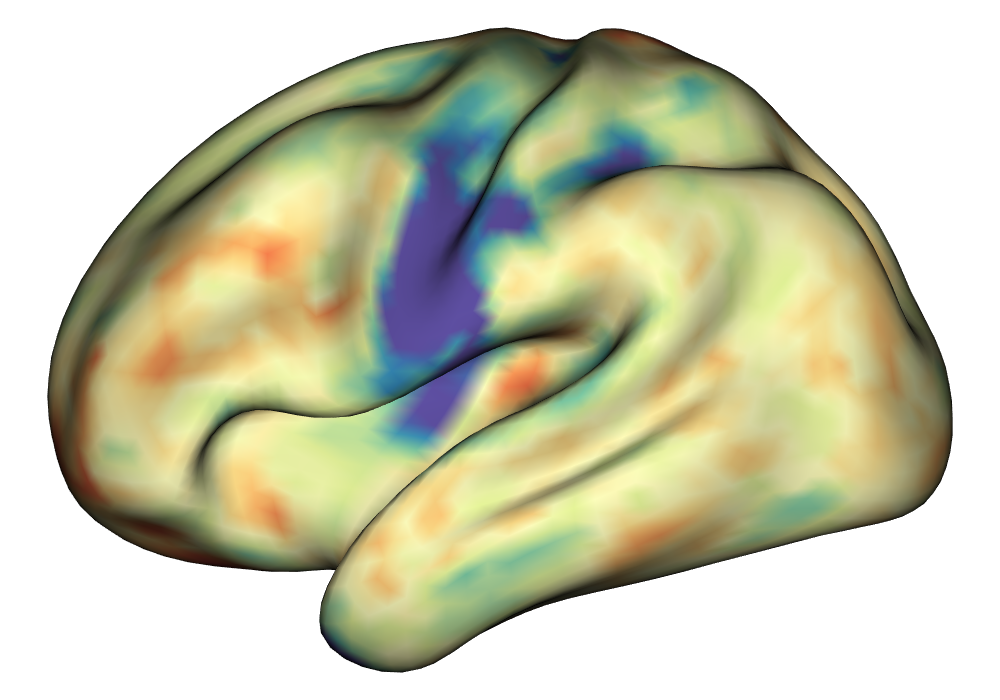} &
    		\Includegraphics[width=0.4\textwidth]{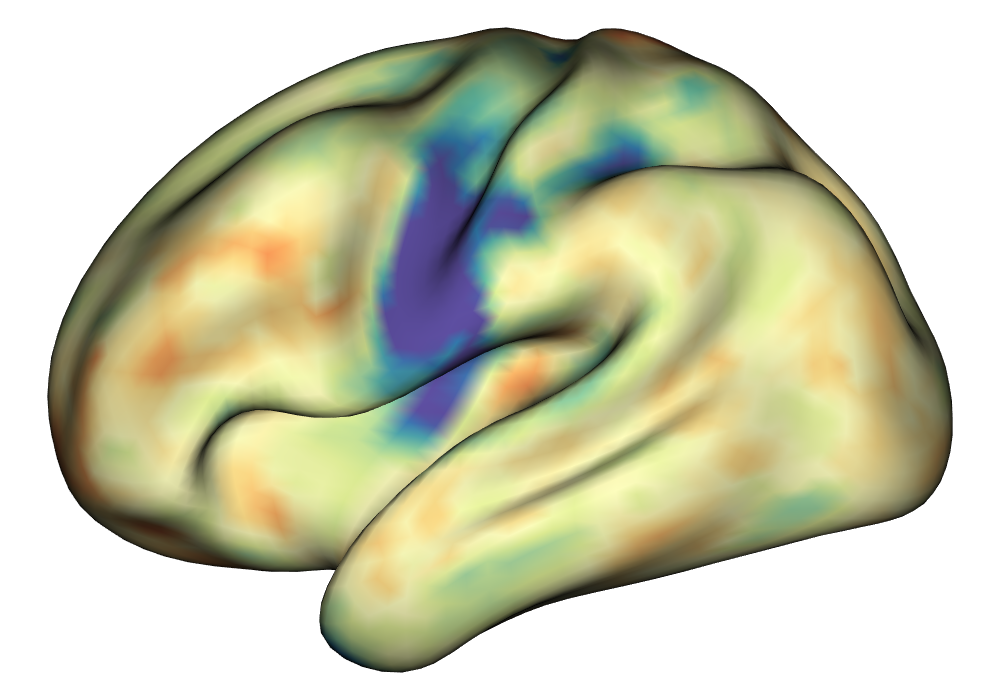} \\
    		\cline{2-3} 
    		\rotatebox[origin=l]{90}{\textbf{Subject B}} &
    		\Includegraphics[width=0.4\textwidth]{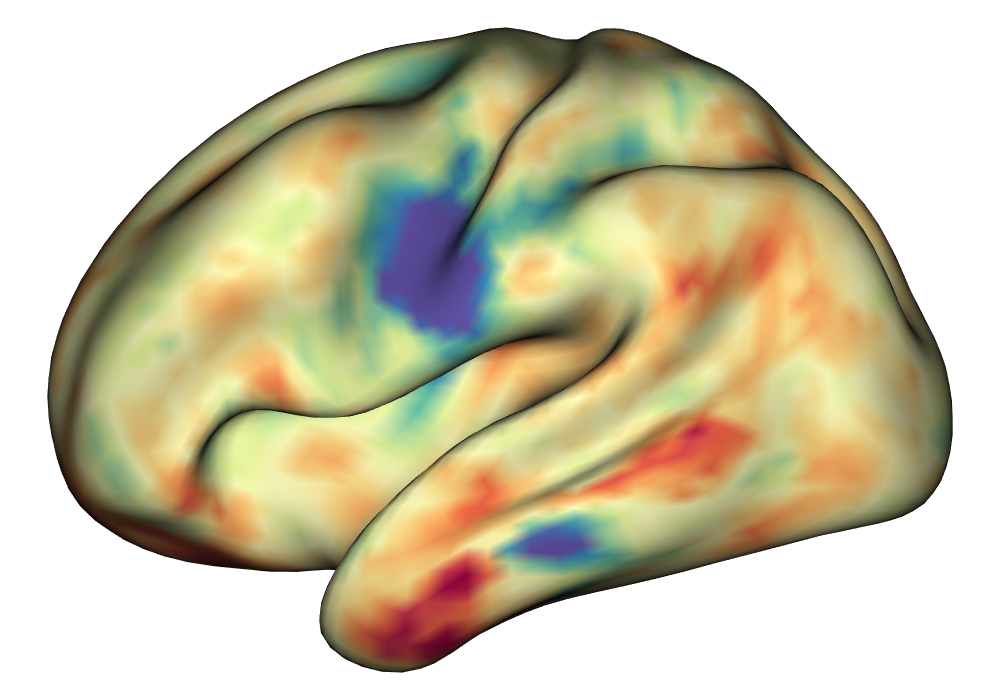} &
    		\Includegraphics[width=0.4\textwidth]{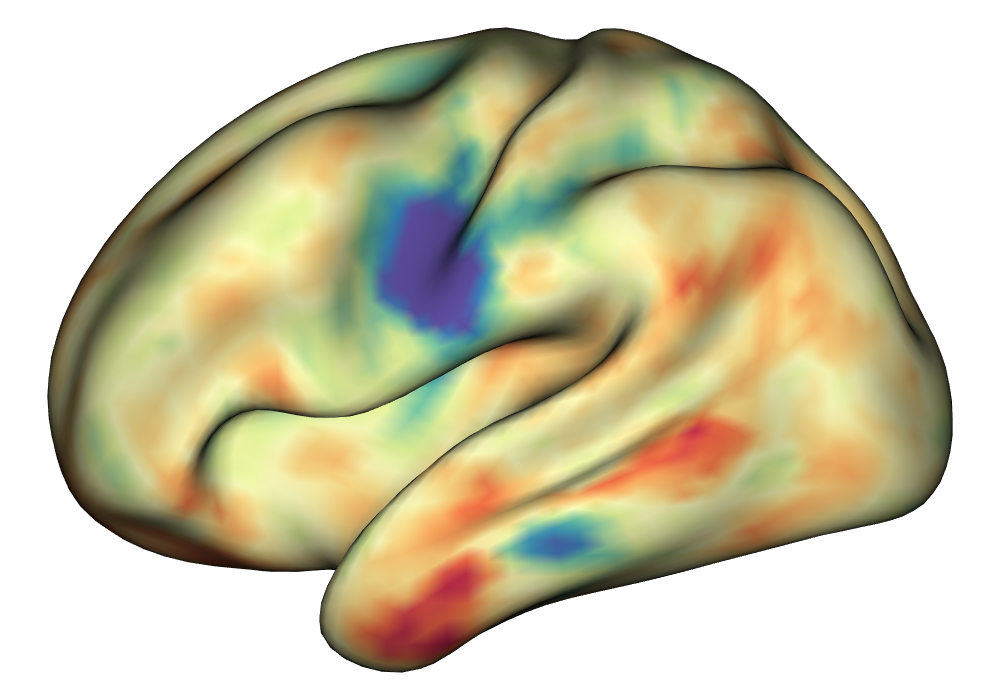} \\
    		\cline{2-3} 
    		\rotatebox[origin=l]{90}{\textbf{Subject C}} &
    		\Includegraphics[width=0.4\textwidth]{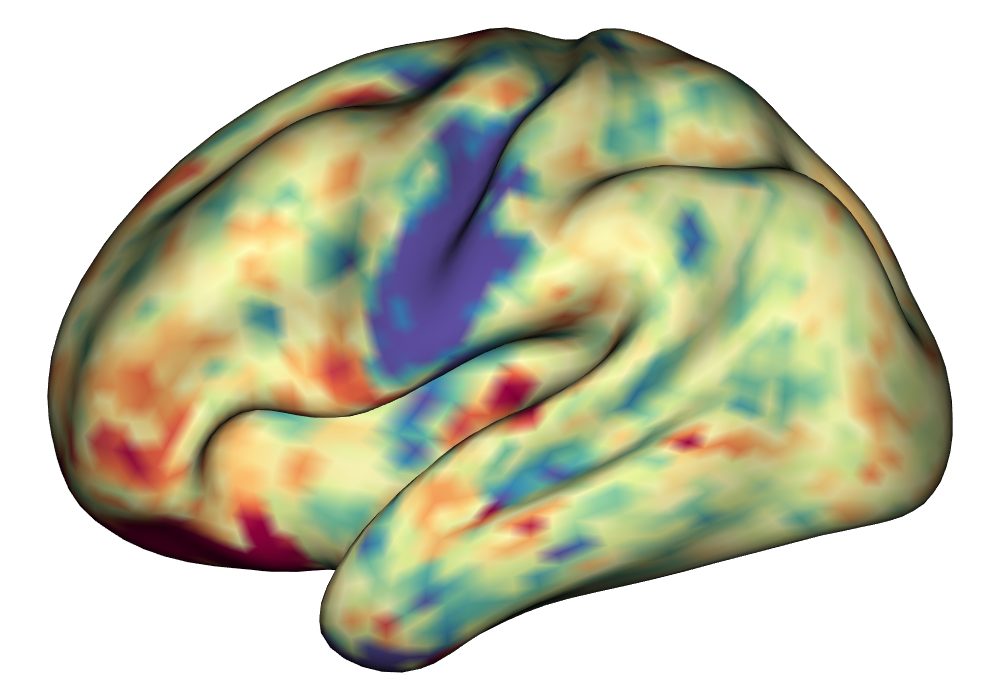} &
    		\Includegraphics[width=0.4\textwidth]{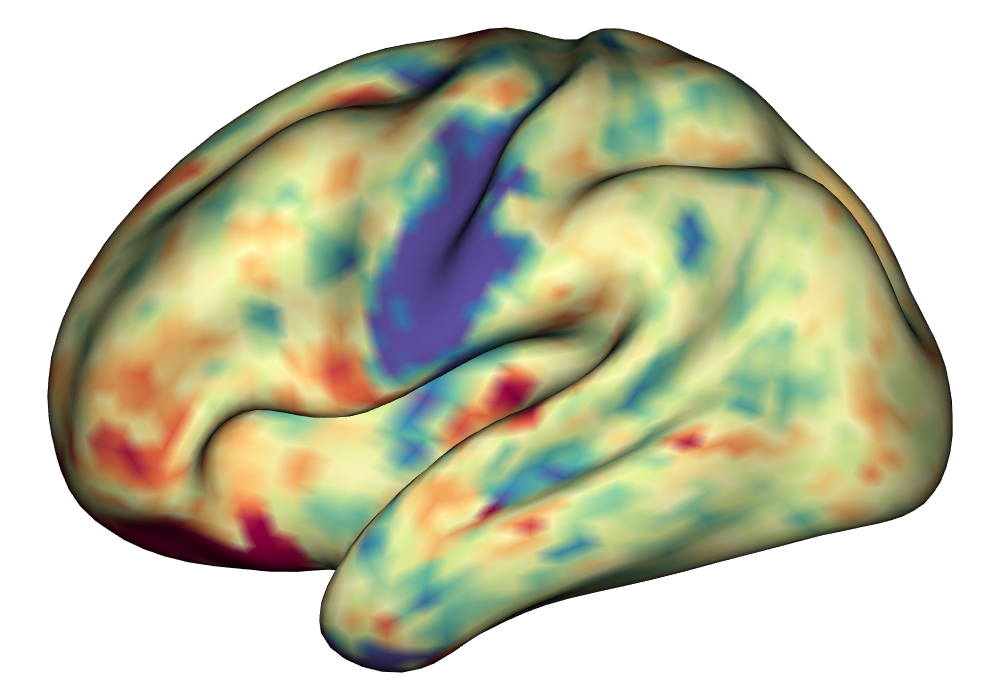} \\
    		\cline{2-3}
    		\multicolumn{1}{c}{\rotatebox[origin=l]{90}{\qquad}} &
    		\multicolumn{2}{c}{\includegraphics[width=0.6\textwidth]{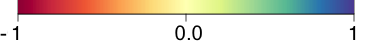}}
        \end{tabularx}
        \caption{Estimates}
    \end{subfigure}
    \begin{subfigure}{0.49\textwidth}
        \begin{tabularx}{\textwidth}{c|c|c|}
            \multicolumn{1}{c}{} &
            \multicolumn{1}{c}{\textbf{INLA}} &
            \multicolumn{1}{c}{\textbf{EM}} \\
            \cline{2-3} 
    		\rotatebox[origin=l]{90}{\textbf{Subject A}} &
    		\Includegraphics[width=0.4\textwidth]{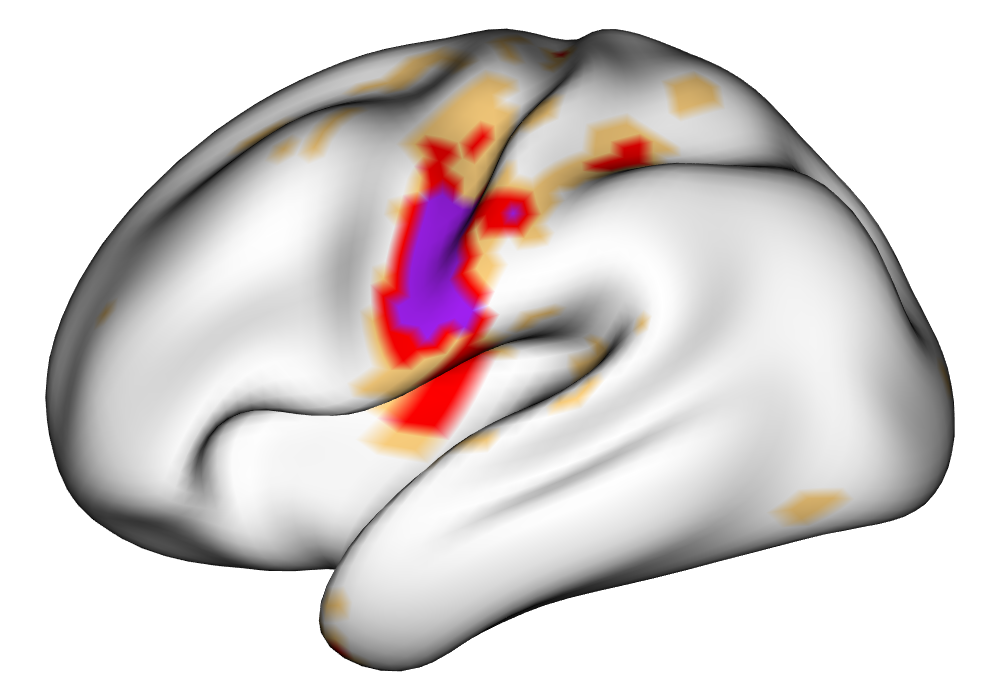} &
    		\Includegraphics[width=0.4\textwidth]{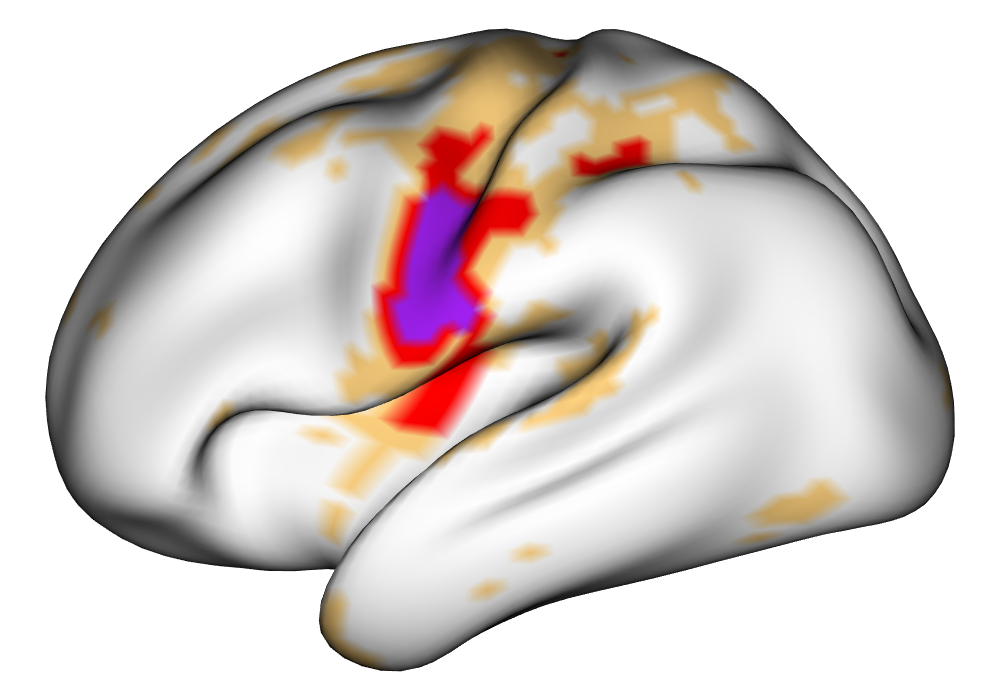} \\
    		\cline{2-3} 
    		\rotatebox[origin=l]{90}{\textbf{Subject B}} &
    		\Includegraphics[width=0.4\textwidth]{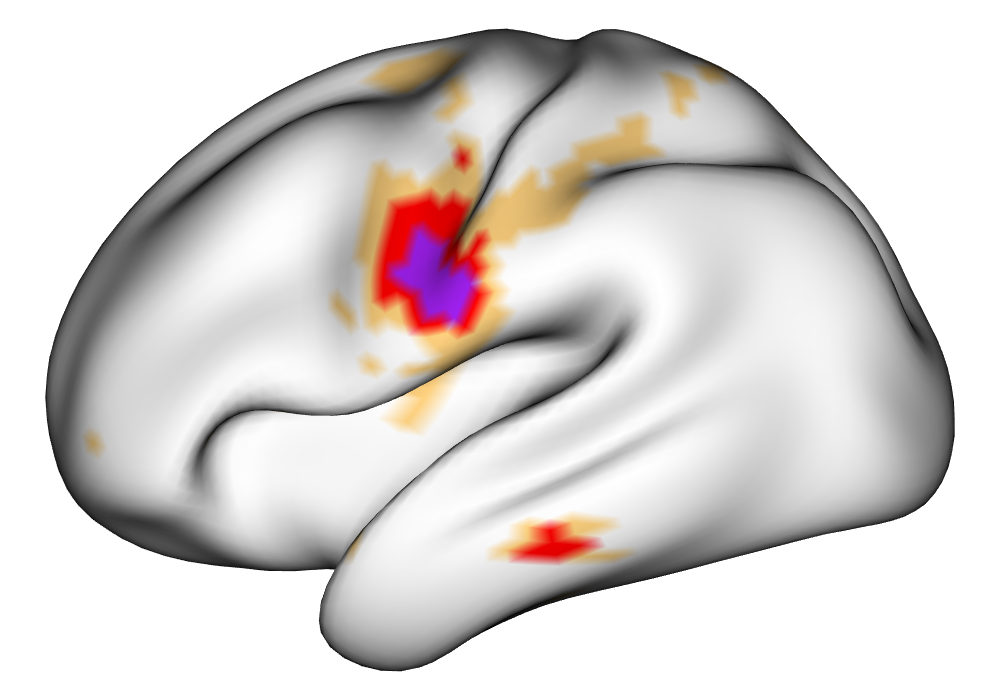} &
    		\Includegraphics[width=0.4\textwidth]{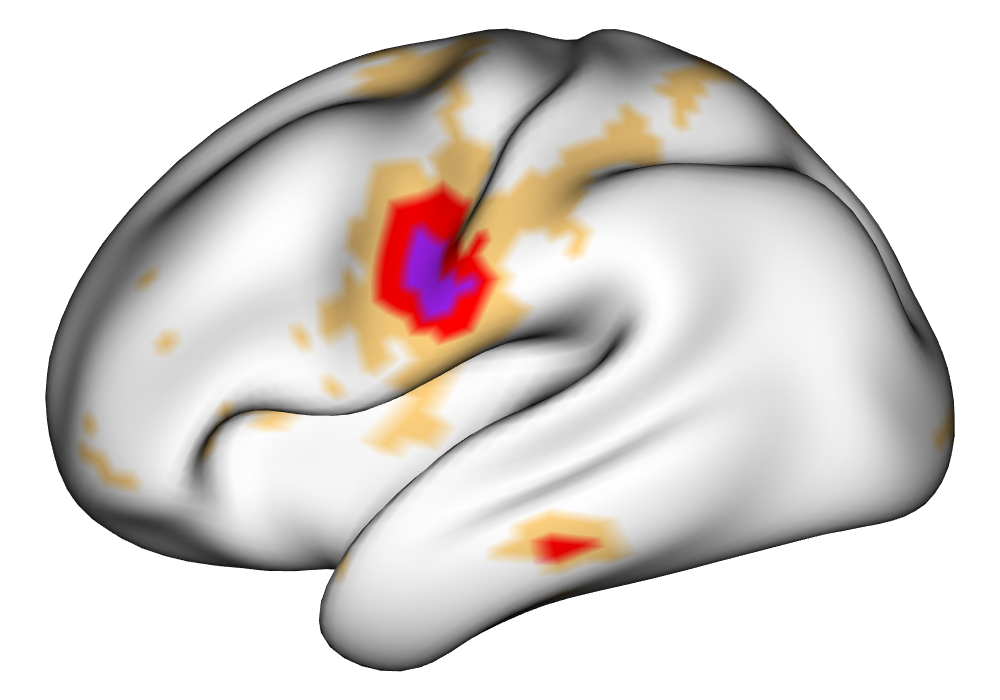} \\
    		\cline{2-3} 
    		\rotatebox[origin=l]{90}{\textbf{Subject C}} &
    		\Includegraphics[width=0.4\textwidth]{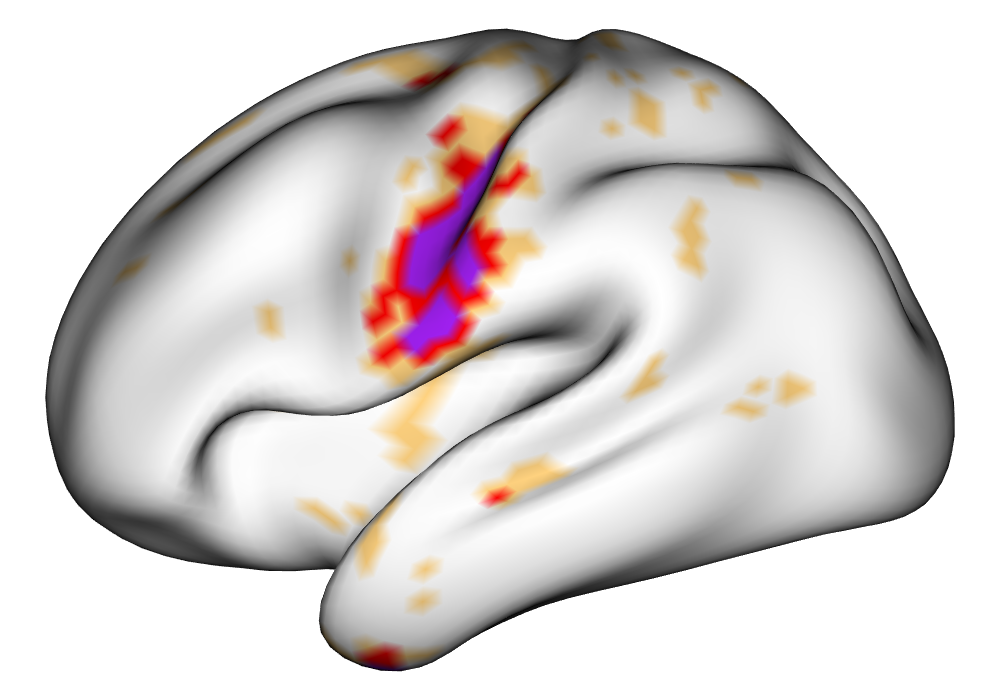} &
    		\Includegraphics[width=0.4\textwidth]{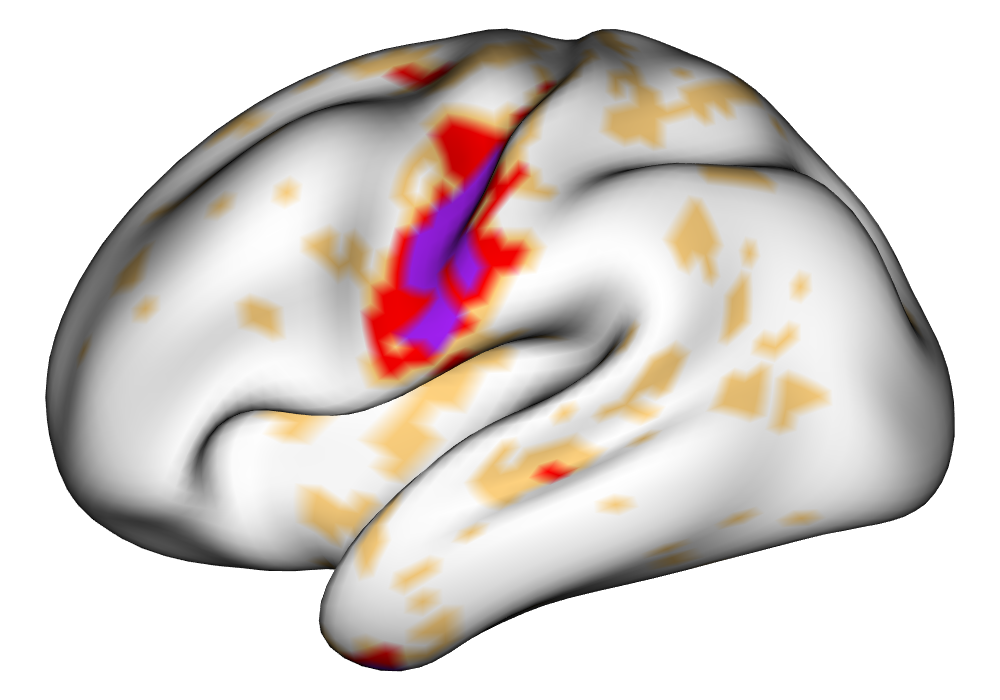} \\
    		\cline{2-3}
    		\multicolumn{1}{c}{\rotatebox[origin=l]{90}{\qquad}} &
    		\multicolumn{2}{c}{$\gamma =$ \textcolor[HTML]{FFD27F}{$\blacksquare$} 0\% 
           \textcolor[HTML]{FF0000}{$\blacksquare$} 0.5\% 
           \textcolor[HTML]{A020F0}{$\blacksquare$} 1\%}
        \end{tabularx}
        \caption{Activations}
    \end{subfigure}
    \caption{Coefficient estimates and activations from the INLA and EM implementations of the Bayesian GLM for the tongue task for three subjects. Estimates are shown with units in \% signal change. Activations are regions determined to be above the threshold $\gamma$ using the excursions method with joint probability of 0.99.}
    \label{fig:hcp_subject_est_and_act}
\end{figure}

\textbf{Figure \ref{fig:hcp_subject_est_and_act}} shows the estimates and the activations found in three different subjects for the tongue task. The tongue task is chosen for display due to the easily-visible pattern of activation in the sensorimotor cortex, which helps to highlight the individual subject differences found in the patterns of estimates and activations. It is clear here that the INLA and EM implementations perform very similarly in terms of the task coefficient estimates, and that the EM implementation shows slightly more activation, particularly at the lowest threshold, $\gamma = 0\%$. As noted in the simulated data study, this is likely due to the underestimation of posterior variance inherent in the EM method due to its treatment of the hyperparameters as fixed.

\begin{figure}
\centering
    \begin{subfigure}{0.6\textwidth}
        \begin{tabularx}{\textwidth}{c|c|c|}
            \multicolumn{1}{c}{} &
            \multicolumn{1}{c}{\textbf{INLA}} &
            \multicolumn{1}{c}{\textbf{EM}} \\
            \cline{2-3} 
    		\rotatebox[origin=l]{90}{\qquad\textbf{Estimates}} &
    		\Includegraphics[width=0.48\textwidth]{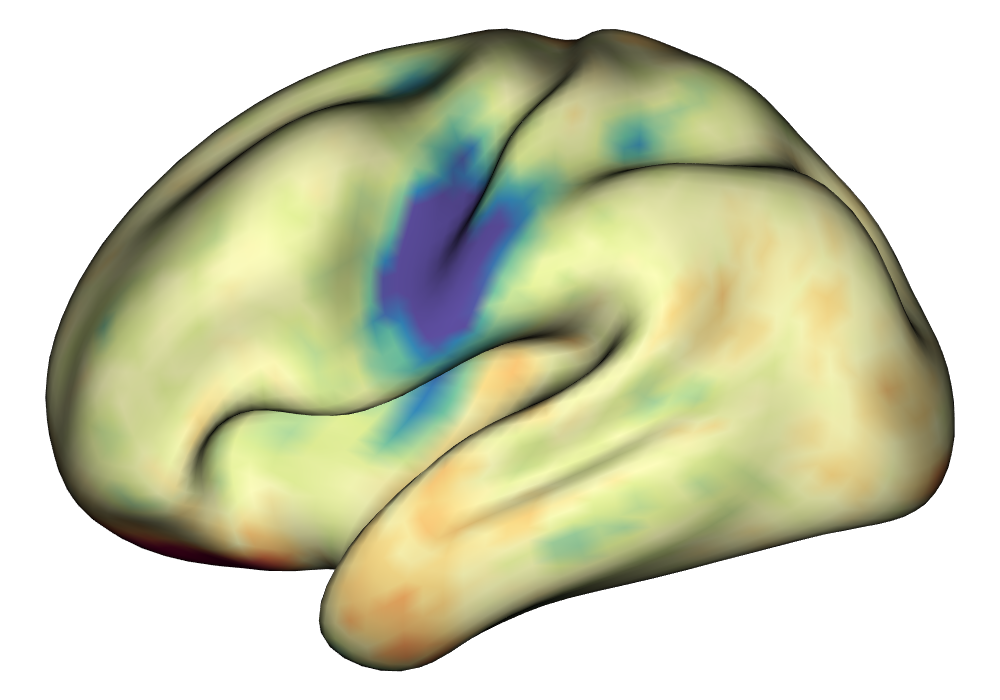} &
    		\Includegraphics[width=0.48\textwidth]{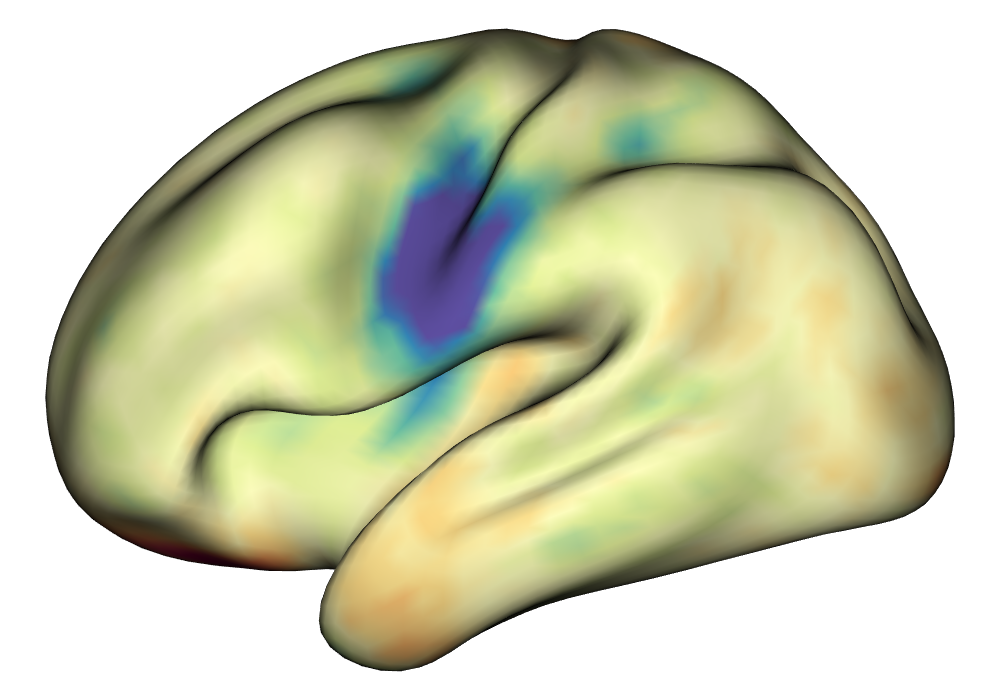} \\
    		\cline{2-3}
    		\multicolumn{1}{c}{\rotatebox[origin=l]{90}{\qquad}} &
    		\multicolumn{2}{c}{\includegraphics[width=0.48\textwidth]{plots/3_legend_spectral_1.png}} \\
    		\cline{2-3} 
    		\rotatebox[origin=l]{90}{\quad \, \textbf{Activations}} &
    		\Includegraphics[width=0.48\textwidth]{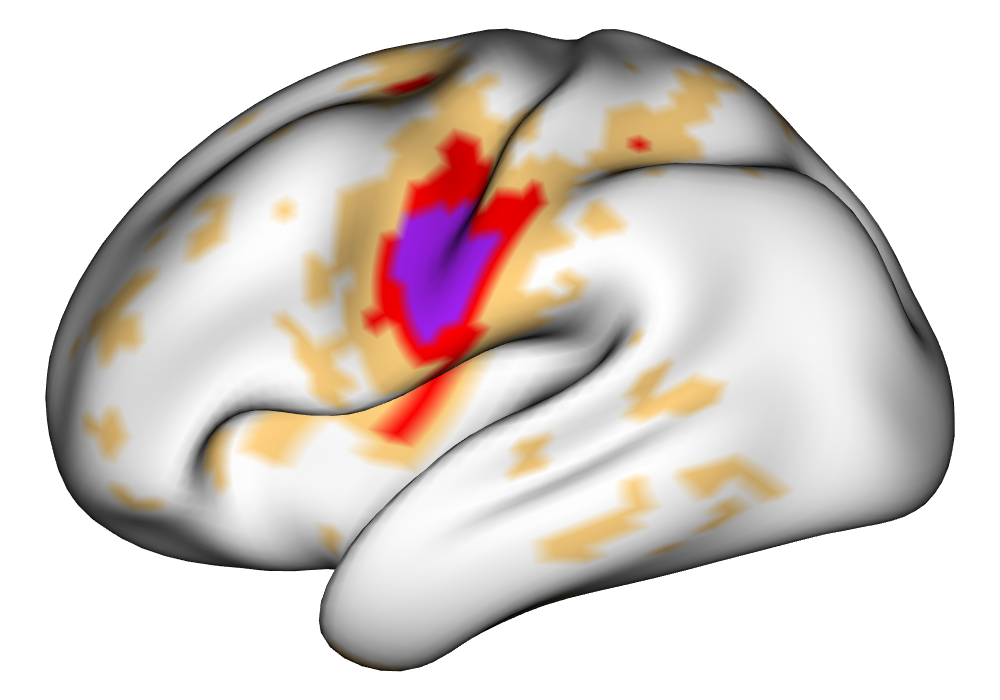} &
    		\Includegraphics[width=0.48\textwidth]{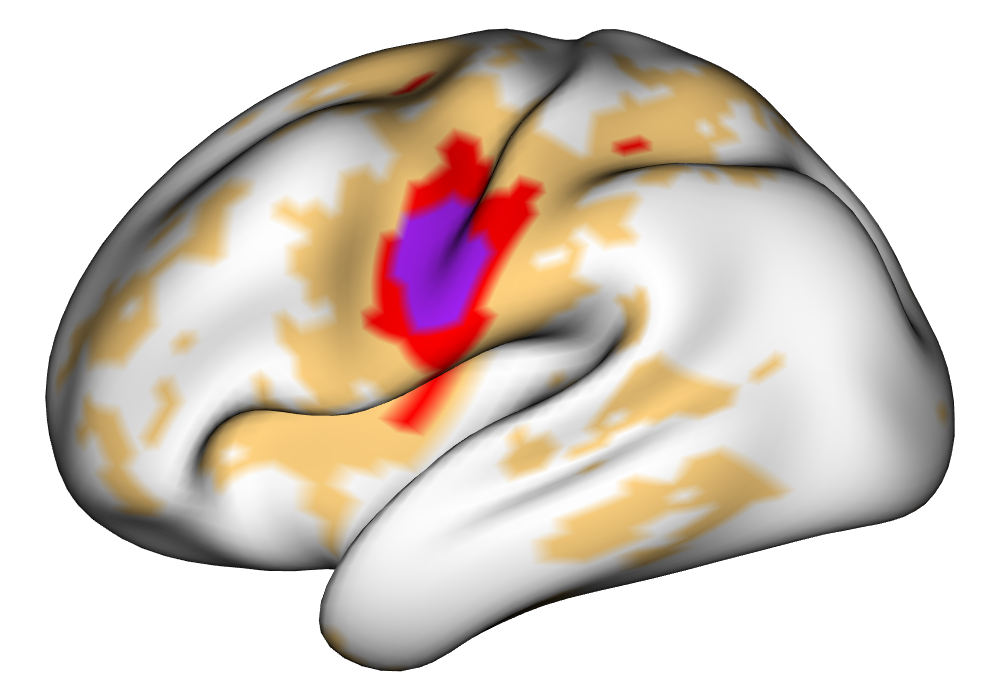} \\
    		\cline{2-3}
    		\multicolumn{1}{c}{\rotatebox[origin=l]{90}{\qquad}} &
    		\multicolumn{2}{c}{$\gamma =$ \textcolor[HTML]{FFD27F}{$\blacksquare$} 0\% 
           \textcolor[HTML]{FF0000}{$\blacksquare$} 0.5\% 
           \textcolor[HTML]{A020F0}{$\blacksquare$} 1\%} 
        \end{tabularx}
    \end{subfigure}
    \caption{Group coefficient estimates and activations from the INLA and EM implementations of the Bayesian GLM for the tongue task for 10 subjects. Estimates are shown with units in \% signal change. Activations are regions determined to be above the threshold $\gamma$ using the excursions method with joint probability of 0.99.}
    \label{fig:hcp_group_est_and_act}
\end{figure}

\textbf{Figure \ref{fig:hcp_group_est_and_act}} shows the group estimates and activations for the tongue task based on 10 subjects. Here we see consistency in the estimates, though the estimates for the EM algorithm bias slightly toward 0 when compared the INLA implementation. Again, the activations show that the EM algorithm detects more activations, especially at the $\gamma = 0\%$ threshold, due to the underestimation of posterior variance stemming from the fixed nature of the hyperparameters. However, the activations found at the $\gamma = 0.5\%$ and $\gamma = 1\%$ thresholds are very similar in appearance, suggesting high levels of agreement at higher, more neurologically meaningful thresholds.

\section{Conclusion}
\label{sec:conclusion}

We developed and implemented an EM algorithm to fit the surface-based spatial Bayesian GLM on cortical surface task fMRI data. This was done in order to reduce the memory consumption required by the INLA implementation of the Bayesian GLM, and also to reduce the dependence on the INLA package. The INLA package, while powerful, is not available on the Comprehensive R Archive Network (CRAN), and may be difficult for investigators without administrator access to install on various computer systems. The increased memory requirement for the INLA package also presents an impediment to using the Bayesian GLM to researchers without significant computing resources at their disposal.

Through analysis of simulated data, we determine that the EM algorithm performs well compared to the INLA implementation, both of which strongly outperform the surface-based classical GLM. 
As expected, the EM implementation of the Bayesian GLM underestimates the posterior variance, inflating the number of active locations detected, though to a much smaller degree when assessing the activations for thresholds above $\gamma = 0\%$.

Analysis of data from the Human Connectome Project motor task data confirms that the EM implementation performs similarly to the INLA implementation, as both methods are able to find sparse, smooth estimates of activation amplitude that display subject-specific characteristics. As in the simulated data, the EM finds more activated locations than INLA at a threshold of $\gamma = 0\%$, but the numbers are very similar for higher thresholds.

Future work on this method includes improving computational efficiency through the use of optimized code. INLA outperforms the EM in smaller data scenarios due to many of its functions being written in C, rather than R. However, exploitation of optimized functions in R and writing other functions in C are expected to greatly increase the speed of the EM implementation of the Bayesian GLM. Forthcoming work in a software paper for the \texttt{BayesfMRI} package in \texttt{R} will make the application of the functions used to perform these analyses accessible to anyone performing inference on relatively modest laptop computers.

\bibliography{BayesGLMEM}
\bibliographystyle{plainnat}

\appendix

\section{Initial value algorithm}
\label{app:initial}

The algorithm for finding the initial values of $\kappa_k$ and $\phi_k$ for task $k = 1,\ldots,K$ is an iterative approach to finding their maximum likelihood estimates. Since the prior on $\mathbf{w}_k$ is
\begin{align*}
    p(\mathbf{w}_k|\kappa_k^2,\phi) & \propto \log |\mathbf{Q}_k|^{1/2} \exp\left\{ -\frac{1}{2} \mathbf{w}_k\mathbf{Q}_k\mathbf{w}_k \right\},
\end{align*}
where $|\mathbf{Q}_k|$ is the determinant of precision matrix $\mathbf{Q}_k$. Taking the log of the prior density and simplifying results in the following estimates for the hyperparameters:
\begin{align*}
    \hat{\phi}_k & = \frac{1}{4\pi n} \mathbf{w}_k'\tilde{\mathbf{Q}}_k \mathbf{w}_k , \\
    \hat{\kappa}_k^2 & = \argmax_{\kappa_k^2} \frac{1}{2} \log |\tilde{\mathbf{Q}}_k| - \frac{1}{8\pi \hat{\phi}_k} \mathbf{w}_k'\tilde{\mathbf{Q}}_k \mathbf{w}_k,
\end{align*}
where $\tilde{\mathbf{Q}}_k = \kappa_k^2\mathbf{C} + 2\mathbf{G} + \kappa_k^{-2} \mathbf{GC}^{-1}\mathbf{G}$. After choosing a reasonable initial value for $\kappa_k^2$ (we choose 4, based on values found using the INLA implementation in previous studies), the algorithm iterates between solving for $\phi_k$ and $\kappa_k^2$ until convergence, which typically happens in 10 to 20 iterations. 

\section{Choice of stopping rule tolerance}
\label{app:epsilon}

\begin{figure}
    \centering
    \includegraphics[width=.7\textwidth]{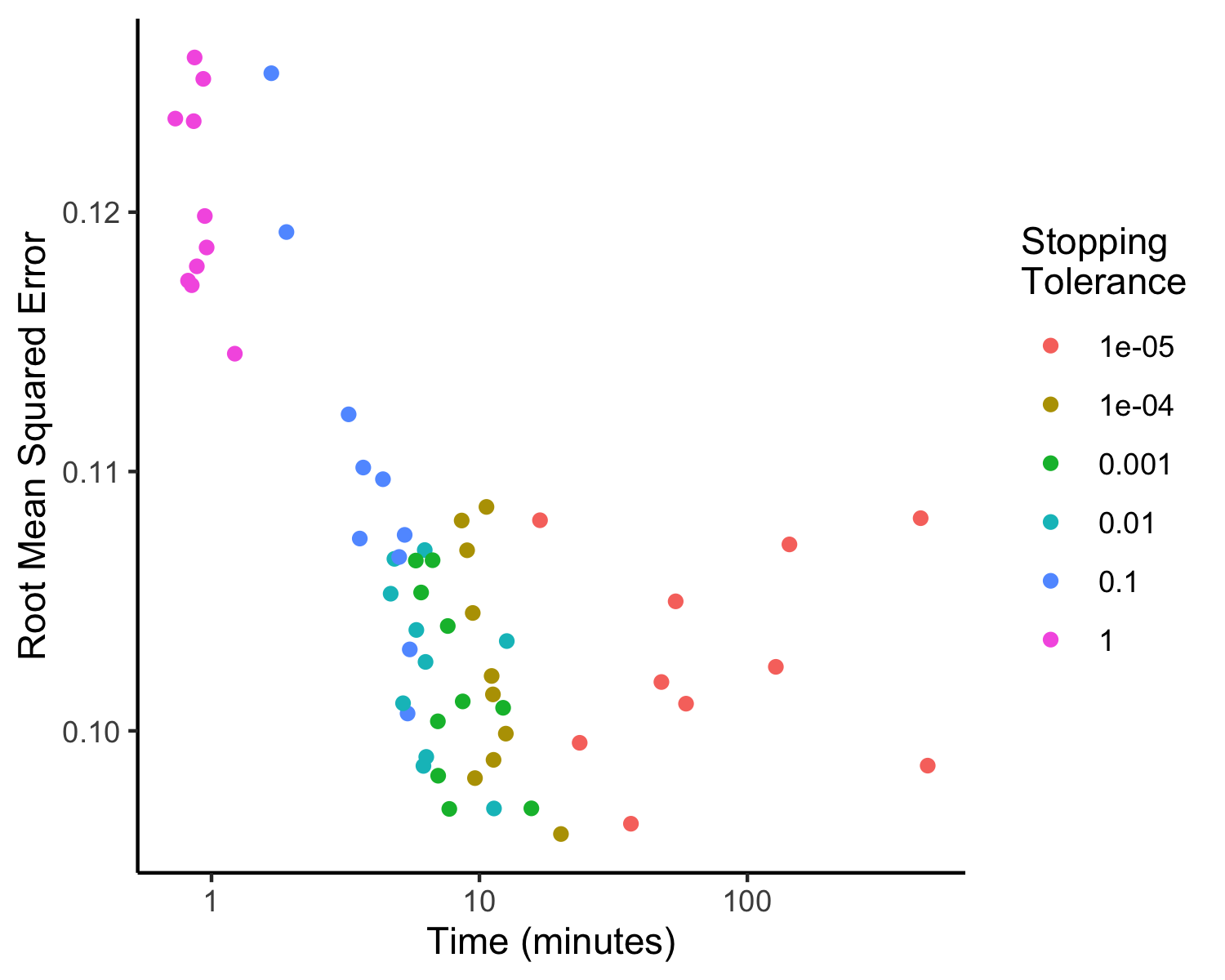}
    \caption{A comparison of computation time and inferential accuracy for different stopping rule tolerances in the EM algorithm.}
    \label{fig:time_rmse_tol}
\end{figure}

A simulation study was performed in which the stopping rule tolerance was allowed to vary from 1 down to 0.001 by powers of 10 to examine the effect of the choice of stopping rule on the speed and accuracy, measured through the square root of the mean squared error (RMSE), of the EM algorithm. In all cases, the model was fitted to simulated cortical surface data on the left hemisphere for four simulated tasks with a spatial resolution of 5,000 vertices per hemisphere. This data generation setting was used to generate 9 different datasets in order to give an idea of the variance in time and  \textbf{Figure \ref{fig:time_rmse_tol}} shows the differences in the accuracy and speed for the different tolerance levels. Based on this analysis, the stopping rule tolerance was set to $\epsilon = 0.001$ for all of the analyses in this paper.

\end{document}